\documentclass[12pt]{article}
\usepackage{amsmath,amsfonts,amssymb}
\usepackage{graphicx}

\newcommand{\be}{\begin{equation}}
\newcommand{\ee}{\end{equation}}
\newcommand{\ba}{\begin{eqnarray}}
\newcommand{\ea}{\end{eqnarray}}
\newcommand{\SO}{\mathrm{SO}}
\newcommand{\SU}{\mathrm{SU}}
\newcommand{\U}{\mathrm{U}}

\usepackage{color}

\begin{document}
\newcommand{\todo}[1]{{\em \small {#1}}\marginpar{$\Longleftarrow$}}   
\newcommand{\labell}[1]{\label{#1}\qquad_{#1}} 


\begin{center} {\Large \bf A strongly coupled zig-zag transition}
\end{center} 

\vspace{7mm}

Vijay Balasubramanian$^{a,b,}$\footnote{\tt email: vijay@physics.upenn.edu}, 
Micha Berkooz$^{c,}$\footnote{\tt email: micha.berkooz@weizmann.ac.il}, 
Simon F. Ross$^{d,e,}$\footnote{\tt email: S.F.Ross@durham.ac.uk}, 
Joan Sim\'on$^{f,}$\footnote{\tt email: j.simon@ed.ac.uk}

\vspace{5mm}

\bigskip\centerline{$^a$\it 
David Rittenhouse Laboratories, University of Pennsylvania}
\smallskip\centerline{\it 209 S 33$^{\rm rd}$ Street, Philadelphia, PA 19104, USA}
\bigskip\centerline{$^b$\it Laboratoire de Physique Th\'{e}orique, \'{E}cole Normale Sup\'{e}rieure,}
\smallskip\centerline{\it 24 rue Lhomond, 75005 Paris, France}
\bigskip\centerline{$^c$\it Department of Particle Physics and Astrophysics, Weizmann Institute of Science,}
\smallskip\centerline{\it  Rehovot 76100, Israel}
\bigskip\centerline{$^d$\it Centre for Particle Theory, Department of Mathematical Sciences} \smallskip\centerline{\it Durham University, South Road, Durham DH1 3LE, UK}
\bigskip\centerline{$^e$\it LPTHE, UPMC, 4 Place Jussieu, 75252 Paris CEDEX 05, France}
\bigskip\centerline{$^f$\it School of Mathematics and Maxwell Institute for Mathematical Sciences}\smallskip\centerline{\it University of Edinburgh, King's Buildings, Edinburgh EH9 3JZ, UK}
\bigskip\medskip

\begin{abstract}
The zig-zag symmetry transition is a phase transition in 1D quantum wires, in which a Wigner lattice of electrons transitions to two staggered lattices. Previous studies model this transition as a Luttinger liquid coupled to a Majorana fermion. The model exhibits interesting RG flows, involving quenching of velocities in subsectors of the theory. We suggest an extension of the model which replaces the Majorana fermion by a more general CFT;  this includes an experimentally realizable case with two Majorana fermions. We analyse the RG flow both in field theory and using AdS/CFT techniques in the large central charge limit of the CFT. The model has a rich phase structure with new qualitative  features, already in the two Majorana fermion case. The AdS/CFT calculation involves considering back reaction in space-time to capture subleading effects. 
\end{abstract}

\section{Introduction}

One-dimensional and quasi-one-dimensional systems are of particular interest in condensed matter physics and possess a rich experimental and theoretical structure. Under favorable circumstances, they remain under theoretical control even when the dynamics is strongly coupled, unlike in higher dimensional systems. A  case of interest is the zig-zag transition, where a one-dimensional electron crystal becomes quasi-one dimensional as we increase its charge density. 
The transition was studied at strong and weak coupling in \cite{cm1,cm2,sitte}. Our aim in this paper is to study generalizations of this model, both in field theory and using a holographic description of the system in a large central charge limit. 

In section \ref{zzc}, we first review the description of the zig-zag system in earlier work, and then discuss our generalisation.  The critical theory at the usual zig-zag phase transition is described by a Luttinger liquid coupled to a single Majorana fermion via a fermion bilinear operator. This exhibits unusual Renormalization Group (RG) behaviour.  Sitte et al. \cite{sitte} found a flow to weak coupling where the velocities $u, v$ of the fermion and the Luttinger liquid flowed to $u/v=1$, $u=v=0$. Our generalisation proceeds in three steps. First, we consider an extension with $k$ Majorana fermions preserving a $\SO(k)_{\text{left}}\times \SO(k)_{\text{right}}$ symmetry.  The case of two fermions is easily achievable experimentally, and it already shows important qualitative changes in the RG behaviour. The flow is no longer generically to $u/v=1$, and $u$ does not flow to zero. Second, for models describing $k >1$ fermions, there is an additional marginal interaction, preserving the above symmetries, with coupling $f_c$.  Physically, if we think of each Majorana fermion as encoding a band which is being filled, the additional coupling describes a charge-charge interaction between bands. We find the set of RG equations, valid for any $k$, for weak coupling between the bands, based only on the $k=1$ results and on symmetry arguments. For all $k >1$, we find that the parameter space is divided into two regions by a critical line at a particular value of $f_c$. The RG flows of the system simplify for $k=2$ and in the limit of large $k$. We solve these RG equations analytically for some special cases, and present numerical results for a typical set of values. The results are summarized in Fig.~\ref{fig:phases}. 

The simplification at large $k$ leads us to consider our third generalisation. It consists of coupling the Luttinger liquid to a generic large central charge Conformal Field Theory (CFT), which contains  an operator ${\cal O}$ with dimension $(1/2,1/2)$.   Integrating out the Luttinger liquid, the resulting theory can be written purely in terms of the CFT with an induced multi-trace, marginal deformation $F\, \mathcal O^2$, where $F$ is a new momentum dependent coupling.

We explain in Sec.~\ref{adscft} that the RG flow of this setup has a natural embedding within the AdS/CFT correspondence; this uses work on multi-trace deformations in \cite{Aharony:2001pa,Witten:2001ua,Berkooz:2002ug} and involves a back-reaction calculation similar to \cite{Gubser:2002zh}. The scalar operator $\mathcal O$ is dual to a scalar field $\Phi$ in the bulk spacetime, and the induced multi-trace deformation determines a boundary condition for the latter. As in the field theory discussion, there is a critical line at a particular value of $F_c$ splitting the RG orbits. On the critical line $v$ flows to zero in the IR; above it $v$ approaches a finite value in the IR; below the critical line, $v$ vanishes at a finite scale. The flow obtained from the AdS/CFT calculation agrees with the large central charge limit of the field theory calculation, and qualitatively agrees with the field theory calculation for a finite number of fermions.  Thus we believe this behaviour is robust and the class of large $c$ models we study may actually belong to the same universality class as $k>2$ Majorana fermions.

In the large $c$ limit,  the flow of $u$ (the CFT velocity) is suppressed.  Thus, to obtain this flow holographically, we need to extend the calculation to subleading order in $c$. In section \ref{br}, we calculate this subleading behaviour by studying the one-loop back-reaction of the bulk scalar $\Phi$ on the spacetime metric. We find that $u$ decreases in the IR, but always flows to a finite value, as in the field theory calculation for $k >1$.

In Sec.~\ref{sec:sum}, we highlight the main results obtained in our analysis. We include two technical appendices at the end of the paper. In Appendix~\ref{tduality} we derive a set of duality transformations which are used to obtain the full set of RG equations in the main text; in Appendix~\ref{angle}, we discuss the computation of the momentum integrals controlling the one-loop contribution to the expectation value of the bulk field $\Phi$ stress tensor.

\section{Generalizing the zig-zag transition}
\label{zzc}

The main goal of our paper is to study generalizations of the critical theory associated with the zig-zag transitions, and the approach to such points. In this section we review the field theory model of the zig-zag transition, and analyse a simple extension.   In Sec.~\ref{zigzaga}, we review the setup and phase diagram of the zig-zag transition. Phrased in the language of two-dimensional CFT, the critical point of the model is a Luttinger liquid coupled to a single $c=1/2$  Majorana fermion (in both the left and right moving sectors), with different velocities for the fermion and the Luttinger liquid. In Sec.~\ref{c2gen}, we generalise to an arbitrary number of Majorana fermions. The generalisation to two fermions may be experimentally realizable. The RG flow in this generalisation is discussed in Sec.~\ref{RGs}. The main point of our discussion is that  the RG flow for more than one fermion is qualitatively different. The model simplifies for a large number of fermions.   This motivates us to introduce a more general model with a Luttinger liquid coupled to a large central charge CFT  in Sec.~\ref{legCmdl}.

\subsection{The zig-zag transition} \label{zigzaga}

The dynamics of electrons in a one dimensional quantum wire is dominated by their Coulomb interactions in the low-density regime $n\,a_B \ll 1$, where $n$ is the electron density and $a_B$ is Bohr's radius in the given material. Experimentally, quantum wires are created by confining 3D electrons to move freely in the $x$ direction by some external potential. If the latter is such that typical excitation energies are low in the $y$ direction and high in the $z$ direction, then motion is only quasi one-dimensional (For a review of the physics of this system see \cite{review}).

Classically, deviations from one dimensional physics arise due to electron lateral motions in the confining potential which can be assumed to be $V_{\text{conf}} = \frac{1}{2}m\Omega^2 \sum_i y_i^2$ ($y_i$ is the transverse coordinate of the $i$'th electron). Thus, physics is parameterised by two tunable parameters: $\Omega$, the frequency of the harmonic oscillations in $V_{\text{conf}}$ and the electron density $n$.

When the density of electrons is low, electrons sit at the bottom of the potential well, $y_i=0$, and form a 1D Wigner crystal. 
When the density increases, electrons find it energetically favourable to form a quasi 1D zig-zag structure, in which the electron follow a pattern $y_i=(-1)^i y_0$. The transition between the phases is the zig-zag transition, which occurs when the electrostatic repulsion energy between the electrons is of the same order as the energy penalty for going up the potential in the $y$ direction, i.e., roughly when
\begin{equation}
V_{\text{conf}}(r_0)=\frac{1}{2}m\Omega^2 y_0^2 \sim V_{\text{int}}=\frac{e^2}{\epsilon r_0}
\end{equation}
where $\epsilon$ is the dielectric constant of the medium. At low energies and independently of the phase, the crystal has acoustic plasmon excitations, i.e. propagating waves of the electron density. At the zig-zag transition, a transverse soft mode appears,  and develops an expectation value above the transition.

The dynamics of the transition was studied at strong and weak coupling in \cite{cm1,cm2,sitte}. 
In \cite{sitte}, an effective model of the physics near this transition (at weak coupling) was obtained by considering the coupling between the effective one-dimensional electronic bands. When we quantize the motion of the electron in the transverse potential well, there are discrete energy levels for motion in the $y$ direction, which descend into different one-dimensional bands, with the lowest band corresponding to electrons sitting in the bottom of the potential.    As the chemical potential is raised, the second band begins to be populated. 

\begin{figure}
\centering
\includegraphics[keepaspectratio,width=0.6\linewidth]{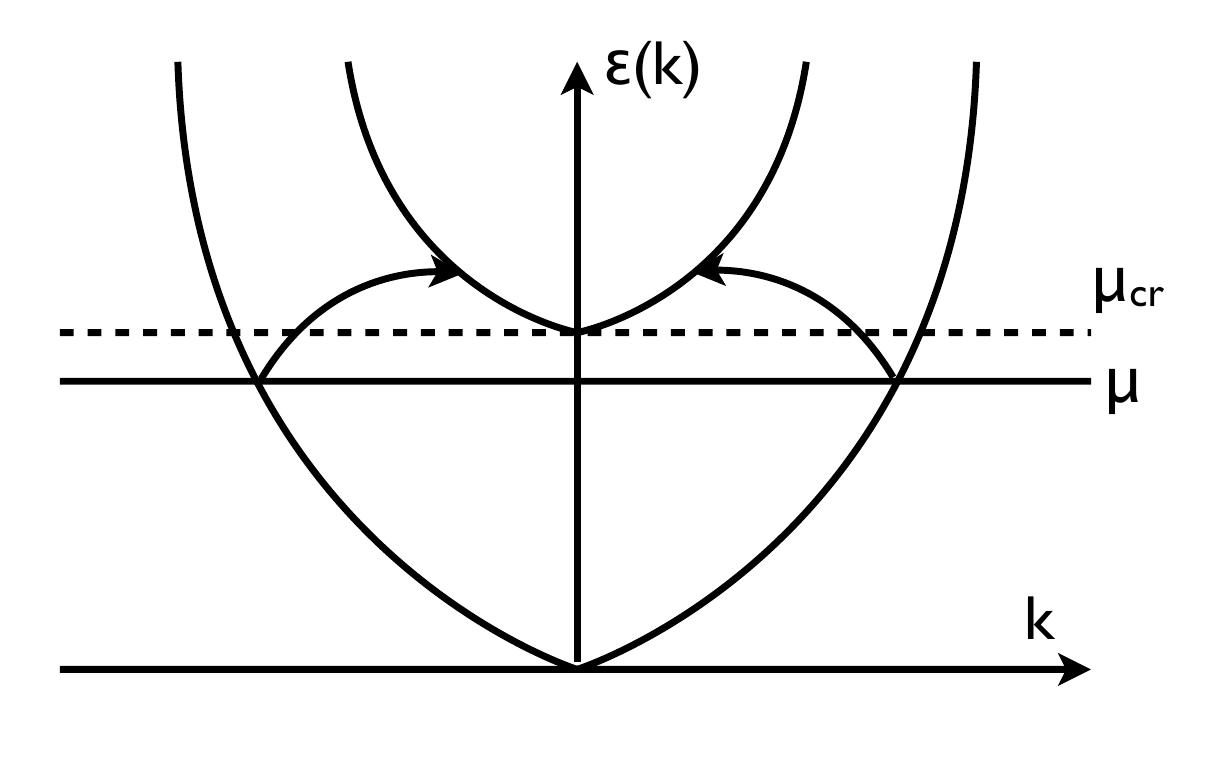}
\caption{Band structure in the zig-zag transition.     The lower band, in which the fermions occupy a straight line configuration, is populated at low chemical potentials.  When the chemical potential increases to $\mu_c$, fermions begin populating the upper band,  in which fermions occupy a zig-zag configuration in two dimensions.
}    
\label{fig:band}
\end{figure}

The near-critical system is modeled by an effective Hamiltonian \cite{sitte}
\begin{equation} \label{hamv1}
{\cal H}_1=\frac{v}{2\pi} \int dx\left(  K (\partial_x \theta) + \frac{1}{K} (\partial_x \phi)^2 \right), 
\end{equation}
\begin{equation} \label{hamv2}
{\cal H}_2 = \int dx \psi^\dagger \left(\frac{\partial_x^2}{2m} -\mu + \mu_{cr} \right) \psi,
\end{equation}
\begin{equation} \label{hamv3}
{\cal H}_{12}=\int dx \left( -\frac{g_x}{\pi} \, \partial_x\phi \,  \psi^\dagger\psi + \frac{u}{2} \left(e^{2i\theta} \,  \psi \partial_x\psi + \text{h.c.}\right)\right).
\end{equation}
${\cal H}_1$ describes the acoustic plasmon excitations in the lowest band, when we fill it up to some Fermi energy, and linearize about it. This is a Luttinger model  written in terms of standard conjugate variables $\theta$ and $\phi$ ($\theta$ is a boson with an $S^1$ target space, and $\phi$ is its dual with an inverse $S^1$ radius, defined by $\partial_x \phi = \partial_t \theta$ up to normalization;  see, e.g., \cite{Senechal:1999us}). ${\cal H}_2$ represents the second band which is just beginning to be populated at the critical chemical potential $\mu_{cr}$, as shown in Fig.~\ref{fig:band}. The interaction between the two bands is encoded in \eqref{hamv3}. The first term describes  electrostatic repulsion between  electrons in the two bands; the second one describes pairs of electrons hopping between the bands.\footnote{To see this, note that the $U(1)$ charge symmetry is implemented as a shift  $\theta\rightarrow \theta + a$ in these variables. Thus, the Noether charge density in the lower band is $\partial_t\theta$. As an operator, the latter is equivalent to $\partial_x \phi$ (up to an overall normalization). Thus $\partial_x\phi\, \psi^\dagger\psi$ is indeed a charge-charge interaction between the bands.  For the second term, note that an  electron (hole) in the lower band is given, in the Luttinger theory, by an operator $e^{i\theta}$ ($e^{-i\theta}$) and hence $e^{2i\theta}$ inserts two electrons into the lower band. These electrons come from the upper band, where they are annihilated using the operator $\psi\partial_x\psi$, which is the lowest (engineering) dimension charge $(-2)$ operator in the upper band (Fig.~\ref{fig:band}).} There is no single electron hopping due to mismatch between the momenta at the Fermi surfaces of the two bands.

It is convenient to rewrite the model by applying a chiral rotation $U=e^{i\int\theta\psi^\dagger\psi}$. After this rotation, it can be described in terms of  a Euclidean Lagrangian density \cite{sitte}
\begin{equation} \label{lagaLL}
{\cal L}_{\text{LL}}={1\over 2\pi vK} \bigl( (\partial_t\phi)^2 + v^2 (\partial_x \phi)^2\bigr) ,
\end{equation}
\begin{equation} \label{ising}
{\cal L}_{\text{Ising}}=\psi^\dagger\partial_t\psi + {u\over2}(\psi \partial_x\psi + \text{h.c.}) + r\psi^\dagger\psi,
\end{equation}
\begin{equation} \label{lagaInt}
{\cal L}_{\text{int}}=-{\lambda\over \pi} \partial_x\phi \psi^\dagger\psi ,
\end{equation}
where $r=\mu_{cr}-\mu$ is the deviation from the critical value of the chemical potential and $\lambda = g_x-\pi v/K$. From now on we will set $r=0$.  The free boson (\ref{lagaLL}) represents the acoustic plasmon excitations. The free fermion $\psi$ in (\ref{ising}) represents the second band.

The RG flow of this model  was calculated in \cite{sitte} at one loop.   The diagrams that contribute are Fig.~\ref{feyn}a,b,c.    Integrating out the modes within the momentum shell $(\Lambda/b,\Lambda)$, \cite{sitte} defined the RG scheme so that $v$ and $K$ are RG invariants.  Since Fig.~\ref{feyn}a is a 1-loop correction to the bosonic propagator, it generates running of the velocity $v$.   We can then make $v$ an RG invariant by rescaling the time and space momenta differently, $k_x \rightarrow k_x /b$ and $k_t \rightarrow k_t /b^z$, where 
\begin{equation} \label{zfact}
z=1+{\lambda^2K\over 4\pi^2 uv}.
\end{equation} 
This value of $z$ is determined by computing the 1-loop correction from the diagram Fig.~\ref{feyn}a.   The RG equations for the remaining parameters are then determined by Fig.~\ref{feyn}b,c to be 
\begin{equation} \label{rgone}
\frac{\partial u}{\partial \log b}=-\frac{\lambda^2K}{\pi^2} \left(
\frac{1}{(v+u)^2} - \frac{1}{4uv}\right)u,
\end{equation}
\begin{equation} \label{rgtwo}
\frac{\partial \lambda}{\partial \log b}=-\frac{\lambda^3K}{2\pi^2} \left(
\frac{1}{u(v+u)} + \frac{1}{(u+v)^2}- \frac{3}{4uv}\right).
\end{equation}
The last term in each of these equations arises from the rescalings of momenta to keep $v$ invariant. The key results are:
\begin{itemize}
\item When $u<v$, the theory flows to $u=v,\ \lambda=0$ in the IR.
\item  When $u=v$, neither $u$ nor $\lambda$ change along the flow. 
\item When $u>v$, the flow is to $u/v,\lambda\rightarrow \infty$, which takes the system outside the perturbative regime. The flow will reach infinite coupling at a finite RG scale.
\end{itemize} 
A drawback of this scheme is that  $u$ and $v$ are not the physical velocities because $k_t$ and $k_x$ are rescaled by different factors.  Sitte et al. \cite{sitte} extracted the physical velocities by multiplying $u$ and $v$ by appropriate factors of scale.  They found that the physical velocities go to zero in the IR.  We will see this more directly in Sec.~\ref{RGs} where we describe the equations in an RG scheme with $z=1$. 
     
\paragraph{RG invariance of $u=v$: } One of the key results of Sitte et al. \cite{sitte} is that the condition $u=v$ is RG invariant, to this order in perturbation theory.    Before considering extensions of this model it is useful to understand where the $u=v$ invariance is coming from.   This feature can be understood as the result of an additional perturbative symmetry of this model with a single Majorana fermion.  
To see the symmetry, note first that the periodicity of $\phi$ is irrelevant to the perturbative calculation, so we can choose it to be such that we can fermionize the Luttinger liquid to two Majorana fermions $\nu_1(z),\nu_2(z), {\bar \nu}_1({\bar z}), {\bar \nu}_2({\bar z})$. In addition $\psi$ can be decomposed into $\nu_3(z)$ and ${\bar \nu}_3({\bar z})$. When $u=v$, the model then manifestly has an $\SO(3)$ symmetry rotating the $\nu_i$, which forms two $\SU(2)$ Kac-Moody algebras at level 2, one for the left and one for the right movers.

We can now classify the operators in the theory in terms of $\SU(2)_L \times \SU(2)_R \times \SO(1,1)$, where the latter is the Lorentz symmetry.\footnote{For $\SU(2)$ we will use the convention that $0$ is a singlet and $1$ a triplet, and for $\SO(1,1)$, we normalise the charges such that the holomorphic stress tensor $T_{zz}$ has charge 2. } 
In particular, the interaction term is 
\begin{equation} \label{nu}
-{\lambda_L\over \pi} \nu_1\nu_2\nu_3{\bar \nu}_3-{\lambda_R\over\pi}{\bar\nu}_1{\bar \nu}_2{\bar \nu}_3 \nu_3,\ \ \ \lambda_L=\lambda_R = \lambda.
\end{equation}
The  first operator in \eqref{nu} has quantum numbers $(0,1,1)$, and the second has quantum numbers $(1,0,-1)$. In general we can  treat $\lambda_{L,R}$  as distinct couplings with opposite quantum numbers to their associated operators.

The operators that deform the velocities are of the form $\nu_i\partial_z \nu_j$, and their anti-holomorphic counterparts. For example, all velocities are changed by the operator $\Sigma_i \nu_i\partial_z \nu_i$, which is the holomorphic stress tensor $T_{zz}$.  The quantum numbers of these operators are either $(2,0,2)$ or $(0,0,2)$ (and their anti-holomorphic counterparts $(0,2,-2)$ and $(0,0,-2)$). Deforming away from $u=v$ is done with the $(2,0,2)$ operator. The coefficient of this operator again has the opposite quantum numbers; assuming it is built out of the couplings $\lambda_{L,R}$, then its quantum numbers tell us that the lowest order in which it can appear is  $\lambda_L^2\lambda_R^4 = \lambda^6$. Thus, this operator can only occur at higher order in perturbation theory, which explains the RG invariance of $u=v$ at one-loop. However, it will generically appear at higher orders,  implying the invariance is only valid to low orders in perturbation theory.   Indeed, we expect that the full RG flow (including all orders in perturbation theory) will not generically approach a Lorentz-invariant theory with $u=v$ in the IR.  When we generalize the model we will also see that the $u=v$ invariance is also special to the case of a single Majorana fermion and fails more generally.

\subsection{Generalising to a larger central charge CFT} 
\label{c2gen}

Instead of a single transverse oscillation, let us consider a generalisation of the previous model describing $k$ transverse oscillations. From a CFT perspective, this corresponds to increasing the number of Majorana fermions and, consequently, the central charge of the theory. The $k=2$ case describes the excitations of electrons moving along the $z$ direction in a shallow confining potential $V=\frac{1}{2} m\Omega^2 (x^2+y^2)$ with rotational symmetry in the two transverse dimensions. In fact, a smaller symmetry group, made out of 90 degree rotations and reflections, i.e., $x\rightarrow -x,\ y\rightarrow -y,\ x\leftrightarrow y$, is sufficient. The $k>1$ generalisations are straightforward, can be experimentally realised (for $k=2$), and introduce qualitative changes in the physics, with $k=2$ as a transition case. 


An effective Lagrangian describing these models is  given by using \eqref{lagaLL} again to describe the plasmon degree of freedom, but replacing \eqref{ising}  by a sum over $k$  Majorana fermions $\psi_i$ $i=1,\dots k$, and choosing the interaction to be:
\begin{equation}
{\cal L}_{\text{int}}=-\lambda\, \partial_x\phi \, \Sigma_{i=1}^k \psi^\dagger_i \psi_i=-\lambda\,\partial_x \phi \, {\cal O}_M,
\label{eq:int}
\end{equation}
where we introduced the notation
\begin{equation}
{\cal O}_M=\Sigma_{i=1}^k \psi^\dagger_i\psi_i. 
\end{equation}
We will assume throughout this article that the $\SO(k)_{\text{left}}\times \SO(k)_{\text{right}}$ supported by the $k$ Majorana fermions is unbroken. 

\begin{figure}
\centering
\includegraphics[keepaspectratio,width=0.9\linewidth]{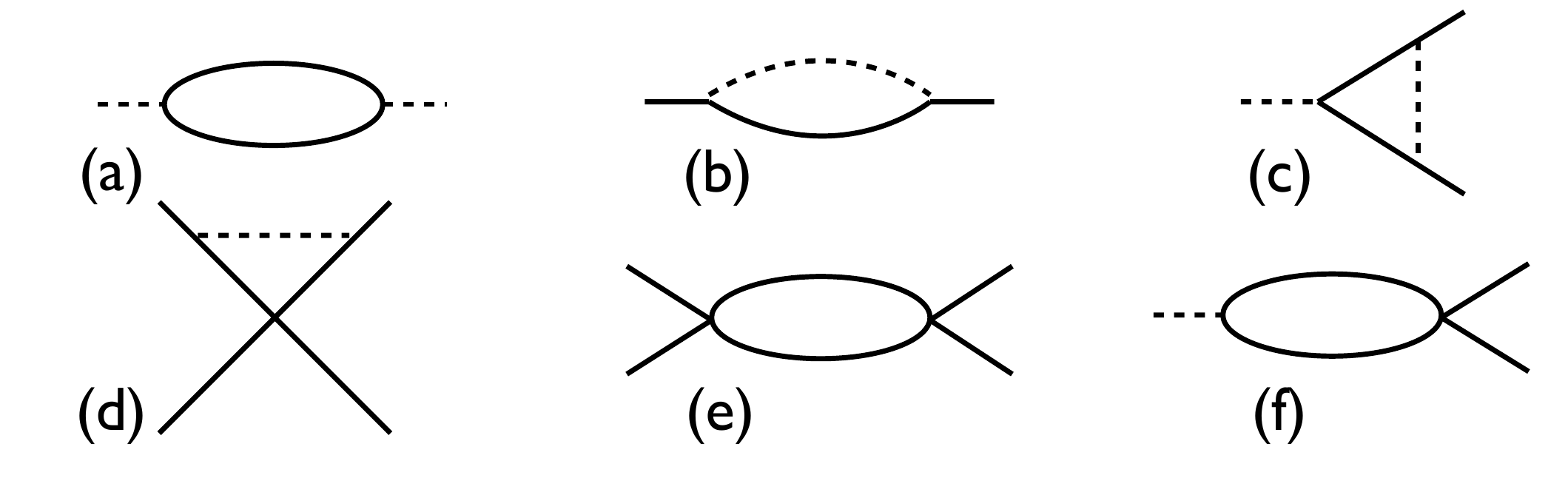}
\caption{Feynman diagrams for processes contributing to the RG flow.  Here the dashed line is the Luttinger scalar and the solid line is one of $k$ fermions.
}    
\label{feyn}
\end{figure}

There are two major distinctions between this case and a single Majorana fermion:
\begin{enumerate}
\item Diagrams with fermion loops are enhanced by a factor of $k$.    In particular, the running of $v$ determined by Fig.~\ref{feyn}a is enhanced in this way relative to the running of $u$.  This will qualitatively change the running we saw before, for example in that $u=v$ will no longer be an RG invariant condition.  Furthermore, if we still wanted to work in the scheme where $v$ is an RG invariant, the value of $z$ would increase with $k$, 
\begin{equation}
z=1+k\frac{\lambda^2K}{4\pi^2 uv}.
\end{equation}
The large value of $z$ (for large $k$) will introduce artificially large $\beta$-functions elsewhere, and hence we will work below either with a $z=1$ scheme or with a scheme where $u$ is an RG invariant. 
\item The Majorana fermion sector contains a new marginal operator for $k>1$. We will denote it as ${\cal O}_M^2$ - a dimension $(1,1)$ operator appearing in the OPE of ${\cal O}_M$ with itself\footnote{Actually, this operator can be written in various other ways. Both left movers and right movers have an $\SO(k)$ symmetry, which is manifest when we decompose $\psi_i=\nu_i(z)+{\bar \nu}_i({\bar z})$, and ${\cal O}_M^2=\Sigma_a J^a(z){\bar J}^a({\bar z})$ where the sum runs over the $N(N-1)/2$ generators of $\SO(N)$. If we bosonize the Majorana fermions into scalars, we obtain new Luttinger liquids and then this operator changes some of the radii of these scalars.}. Since ${\cal O}_M$ is the charge density in the higher bands, ${\cal O}_M^2$ describes the charge interaction within these bands. This is a natural marginal interaction that we could add to the effective interaction lagrangian
\begin{equation}
  {\cal L}_{int} = -\lambda\,\partial_x \phi\, {\cal O}_M + f_c\, \mathcal O_M^2\,.
\label{eq:newint}
\end{equation}
As before, $\partial_x\phi \, {\cal O}_M$ describes the charge interaction between the Luttinger liquid electrons and the electrons in the higher bands. 
\end{enumerate}
We will study below how the addition of the $f_c$ coupling changes the RG flow.

\subsection{RG equations in different schemes} \label{RGs}

The RG equations \eqref{rgone}-\eqref{rgtwo} in \cite{sitte} were derived in a $v$ fixed scheme and scaling $k_t$ and $k_x$ differently. In this section, we will discuss these equations in an RG scheme where we either fix $u$ (the velocity of the fermion) or set $z=1$. The first scheme is natural at large $k$, since the propagator of the Luttinger scalar is strongly renormalized by the diagram Fig.~\ref{feyn}a  which is proportional to $k$, whereas the renormalization of the fermion propagators and the interaction terms (Figs.~\ref{feyn}b and  Fig.~\ref{feyn}c) are $k$ independent. Thus, the velocity of the scalar changes significantly with RG flow, whereas that of the fermions does not. Physically, the scalar is dragged by many fermions whereas each fermion is only dragged by a single scalar. A more natural prescription is therefore to keep the velocity of the fermions fixed. The second scheme $(z=1)$ is always natural since it deals with physical velocities and couplings.

In this section, we discuss the RG equations in these schemes. First,  we discuss the $z=1$ scheme for the $k=1$ scenario and then  generalize to arbitrary $k$.  Afterwards, we discuss the fixed $u$ scheme for $k>1$.

\paragraph{$z=1$ RG scheme:}   Recall that in the scheme where $v$ is an RG invaraint, the momenta are scaled as $k_x \to k_x/b$ and $k_t \to k_t/b^z$.    To go to a scheme where these momenta are scaled in the same way, we undo the above scaling by a rescaling $t \to t b^{-(z-1)}$ as each momentum shell is integrated out.   We can therefore convert the RG equations  \eqref{rgone}-\eqref{rgtwo} obtained in \cite{sitte} to the $z=1$ scheme by rescaling $t$ with $z$ as in \eqref{zfact}.   Requiring the invariance of the Lagrangian \eqref{lagaLL}-\eqref{lagaInt} then implies the rescaling of the couplings $u \to u b^{(z-1)}$, $v \to v b^{(z-1)}$, $\lambda \to \lambda b^{(z-1)}$. When $v$ varies, it is also more convenient to canonically normalise the Lagrangian, absorbing the factor of $(2 \pi v K)^{-1}$ in \eqref{lagaLL} by rescaling the scalar field $\phi$. This produces a redefinition of the coupling
\begin{equation}
  \bar \lambda = \lambda \sqrt{2 \pi v K}\,.
\end{equation}
Using this redefinition, the RG equations in the $z=1$ scheme are
\begin{equation} \label{1rg1}
 \frac{\partial v^2}{\partial \log b} = - {\bar \lambda^2 \over 4\pi^3 u},
\end{equation}
 \begin{equation} \label{1rg2}
 \frac{\partial u}{\partial \log b}=-\frac{\bar \lambda^2 u}{2 \pi^3 v (u+v)^2},
\end{equation}
\begin{equation} \label{1rg3}
\frac{\partial \bar \lambda}{\partial \log b}=-\frac{\bar \lambda^3}{4\pi^3 v} \left(
\frac{1}{u(v+u)} + \frac{1}{(u+v)^2} \right). 
\end{equation}
Since $u, v$ and $\bar \lambda$ are the physical velocities and coupling, we can immediately see that all of them decrease along the RG flows. This is consistent with the result of \cite{sitte} that the physical velocities both go to zero for the RG flows with $u <v$. We also see that there is a rescaling of the physical coupling which was not stressed in \cite{sitte}. Although the coupling $\lambda$ in the fixed $v$ scheme increases for $u > v$, the physical coupling actually decreases. The pathology of the $u >v$ flows is really that they reach $v=0$ at a finite RG scale. We can see from the above equations that this is actually the only possible pathology in the physical couplings; both $u=0$ and $\bar \lambda =0$ are RG invariant conditions, so as all parameters are decreasing, $u$ and $\bar \lambda$ will either approach constant values or asymptotically approach zero in the IR. But since $v=0$ is not an RG invariant condition, $v$ can reach zero at a finite RG scale.  

The generalisation of the RG equations in the $z=1$ scheme to $k>1$ contains two effects:
\begin{enumerate}
\item The one-loop renormalisation of the bosonic propagator picks up a factor of $k$, for the $k$ different species of fermions running in the loop. Since this factor only contributes to the running of $v$, the only modification is to multiply the RHS of \eqref{1rg1} by $k$.
\item Due to the new interaction \eqref{eq:newint}, there will be three new one-loop diagrams; one contribution to the running of the $\lambda$ coupling from this new interaction (Fig.~\ref{feyn}f), and two contributions to the running of $f_c$ (Fig.~\ref{feyn}d,e). 
\end{enumerate}
The resulting $z=1$ scheme RG equations are 
\begin{equation} \label{krg1}
\frac{\partial v^2}{\partial \log b} = - k{\bar \lambda^2 \over 4\pi^3 u},
\end{equation}
 \begin{equation} \label{krg2}
 \frac{\partial u}{\partial \log b}=-\frac{\bar \lambda^2 u }{2 \pi^3 v (u+v)^2},
\end{equation}
\begin{equation} \label{krg3} 
\frac{\partial \bar \lambda}{\partial \log b}=-\frac{\bar \lambda^3}{4\pi^3 v} \left(
\frac{1}{u(v+u)} + \frac{1}{(u+v)^2}\right) - \frac{f_c \bar \lambda (k-1)}{4 \pi^3 u}.   
\end{equation}
\begin{equation} \label{krg4} 
\frac{\partial f_c}{\partial \log b} =  - \frac{f_c \bar \lambda^2}{2\pi^3 v} \left(
\frac{1}{u(v+u)} + \frac{1}{(u+v)^2}\right) - \frac{ f_c^2 (k-2)}{4 \pi^3 u}.
\end{equation}
The factor of $k-1$ in \eqref{krg3} originates from the fermion in the loop in figure~\ref{feyn}f, which can be any species other than the external one. Similarly, the factor of $k-2$ in \eqref{krg4} comes from the fermion loop in figure~\ref{feyn}e, which can be any species other than the two external ones. As for $k=1$, all the right hand sides are non-positive. Thus, all physical parameters decrease along the flow, unless $u =0$, $\bar \lambda =0$, or $f_c=0$, which are RG invariant conditions. Since $v=0$ is not an RG invariant condition, RG flows can reach $v=0$ at finite scale. 

\paragraph{Derivation of the RG equations:} These RG equations can be obtained by explicitly evaluating the one-loop diagrams, but it is useful to note that their form is essentially determined from the  equations at $f_c =0$ by exploiting a symmetry of the system, which involves a transformation between the description in terms of the boson $\phi$ and its dual $\theta$ (T-duality in the particle physics language).  As discussed in Appendix~\ref{tduality},  the change between $\theta$ and $\phi$ involves a shift of the $\mathcal O_M^2$ coupling; that is,  $i \partial_t  \theta \mathcal O_M$ is related to $\partial_x \phi \,  \mathcal O_M + \alpha \mathcal O_M^2$. The coefficient $\alpha$ can be worked out  by using the T-duality rules (see Appendix~\ref{tduality}).   The appropriate coefficient can also be obtained by considering the process of integrating out the boson  in the path integral. We work in terms of the original unrescaled fields, where the Euclidean Lagrangian for $\phi$ is \eqref{lagaLL}. If we assume we have the interaction $\lambda  \, \partial_x \phi \,  \mathcal O_M + \alpha \, \mathcal O_M^2$, integrating out $\phi$ gives 
\begin{equation} \label{dtphi}
\left[ - 2 \pi v K \frac{\lambda^2 k_x^2}{k_t^2 + v^2 k_x^2} +\alpha \right]  \mathcal O_M^2.
\end{equation}
In terms of the dual $\theta$ variable, the Euclidean Lagrangian is $\mathcal L_{LL} = \frac{K}{2 \pi v} ( (\partial_t \theta)^2 + v^2 (\partial_x \theta)^2)$, and we assume that the interaction in terms of this variable is simply $\lambda_t \, \partial_t \theta \,  \mathcal O_M$. Integrating out $\theta$ we will find an effective interaction 
\begin{equation} \label{dttheta}
-\frac{2 \pi v}{K} \frac{\lambda_t^2 k_t^2}{k_t^2 + v^2 k_x^2} \mathcal O_M^2. 
\end{equation}
Thus requiring that the effective interactions agree we  see that $\lambda_t = i  \lambda K/v $ and $\alpha = 2 \pi K \lambda^2/v =  \bar \lambda^2/v^2$. 

Furthermore, we can relate the interaction $\partial_t \theta \mathcal O_M$ back to an interaction of the form $\partial_x \theta \mathcal O_M$ by exploiting the fact that we are working in Euclidean space and can interchange the $t$ and $x$ coordinates. The Lagrangian \eqref{lagaLL} will be symmetric under interchanging the $t$ and $x$ coordinates, if we invert the velocities $u \to 1/u$, $v \to 1/v$. This interchange involves a scaling of the fermions, so in \eqref{lagaInt} we must also rescale $\lambda_x = \lambda_t/u$. Finally, to eliminate the overall factor of $K$ difference between the Lagrangian in terms of $\theta$ and in terms of $\phi$ we should rescale the fields. In total then, using the transformation to the dual variable $\theta$ predicts that the theory has a symmetry under the redefinition of the parameters 
\begin{equation}
\bar \lambda \to i  \frac{\bar \lambda}{uv^2}, \quad u \to \frac{1}{u}, \quad v \to \frac{1}{v}, \quad f_c \to \frac{1}{u^2} \left(f_c - \frac{\bar \lambda^2}{v^2}\right). 
\end{equation}
It is easy to check that \eqref{krg1}-\eqref{krg4} are indeed invariant under this automorphism. (For $k=1$, $f_c$ is not defined, and \eqref{krg1}-\eqref{krg3} are invariant under the action of the transformation on $u, v, \bar \lambda$.) This enables us to determine the values of the coefficients of the $f_c$ terms in \eqref{krg3} and \eqref{krg4} without explicit calculation. Note also that this transformation maps $f_c=0$ to $f_c = \bar \lambda^2/v^2$ and vice-versa. Thus, it predicts that $f_c = \bar \lambda^2 /v^2$ is an RG invariant condition. This can be directly verified from \eqref{krg1}-\eqref{krg4}
\begin{equation}
  \frac{\partial \left(f_c - \frac{\bar\lambda^2}{v^2}\right)}{\partial\log b} = \frac{\left(f_c - \frac{\bar\lambda^2}{v^2}\right)}{4\pi^3uv(u+v)^3}\left(v(u+v)^2\left(2f_c-k\left(f_c-\frac{\bar\lambda^2}{v^2}\right)\right)-2\bar\lambda^2(2u+v)\right)\,.
\label{RGdiv}
\end{equation}

\paragraph{Fixed $u$ RG scheme:} The $z=1$ RG scheme is conceptually simpler, but to perform explicit calculations, it is more convenient to work in a scheme where $u$ is fixed. This can be  obtained by a rescaling of $t$ in a way similar to that described above for converting between the fixed $v$ and $z=1$ schemes.  The resulting value of $z$ is
 \begin{equation} \label{uz}
z=1+\frac{\bar \lambda^2}{2 \pi^3 v (u+v)^2},
\end{equation}
and the RG equations are 
\begin{equation} \label{ukrg1}
\frac{\partial v^2}{\partial \log b} = - {\bar \lambda^2 \over \pi^3} \left( \frac{k}{4u} - \frac{v}{(u+v)^2} \right) = - {\bar \lambda^2 \over \pi^3} \frac{(ku^2 + (2k-4) uv + kv^2)}{4u (u+v)^2},
\end{equation}
\begin{equation} \label{ukrg3} 
\frac{\partial \bar \lambda}{\partial \log b}=-\frac{\bar \lambda^3}{4\pi^3 v} \left(
\frac{1}{u(v+u)} - \frac{2}{(u+v)^2}\right) - \frac{f_c \bar \lambda (k-1)}{4 \pi^3 u},
\end{equation}
\begin{equation} \label{ukrg4}
\frac{\partial f_c}{\partial \log b} =  - \frac{f_c \bar \lambda^2}{2\pi^3 v} 
\frac{1}{u(v+u)}  - \frac{ f_c^2 (k-2)}{4 \pi^3 u}.
\end{equation}
This fixed $u$ scheme is convenient for understanding the relative flow of $u$ and $v$. We can see from \eqref{ukrg1} that $u=v$ is an RG invariant condition for $k=1$, as shown in \cite{sitte}, but for $k>1$ the RHS is negative. Thus, $v$ always decreases relative to $u$ for $k>1$. In fact, $u=v$ defines a line of fixed points of the fixed $u$ RG flow for $k=1$, as in \cite{sitte}, because the first term in \eqref{ukrg3} vanishes. For $k>1$, the second term generates a flow to small $\bar \lambda$ along the $u=v$ line. 

\subsubsection{Solving the RG equations}

\begin{figure}
\centering
\includegraphics[width=0.328 \textwidth]{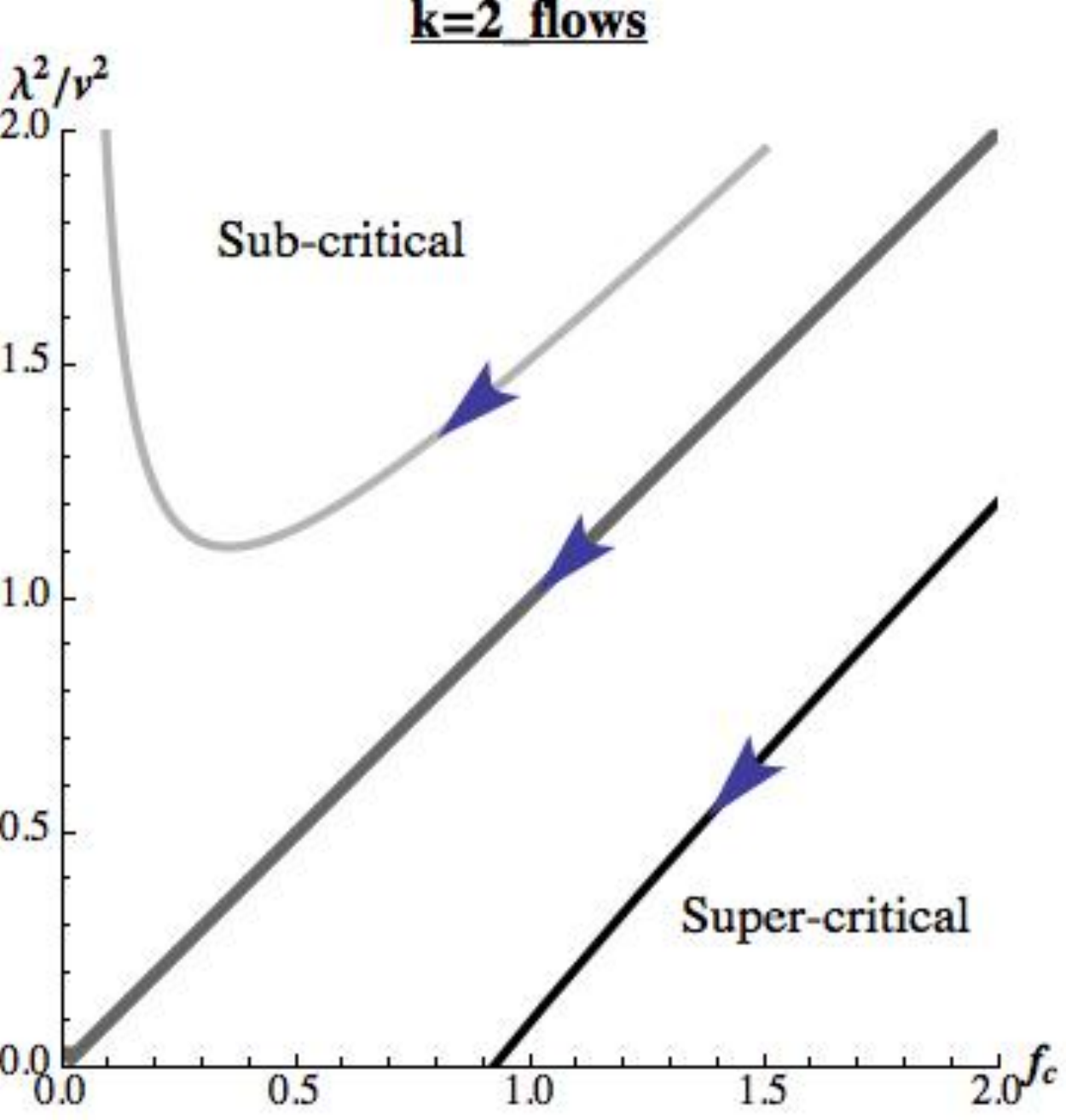} \hfil 
\includegraphics[width=0.328 \textwidth]{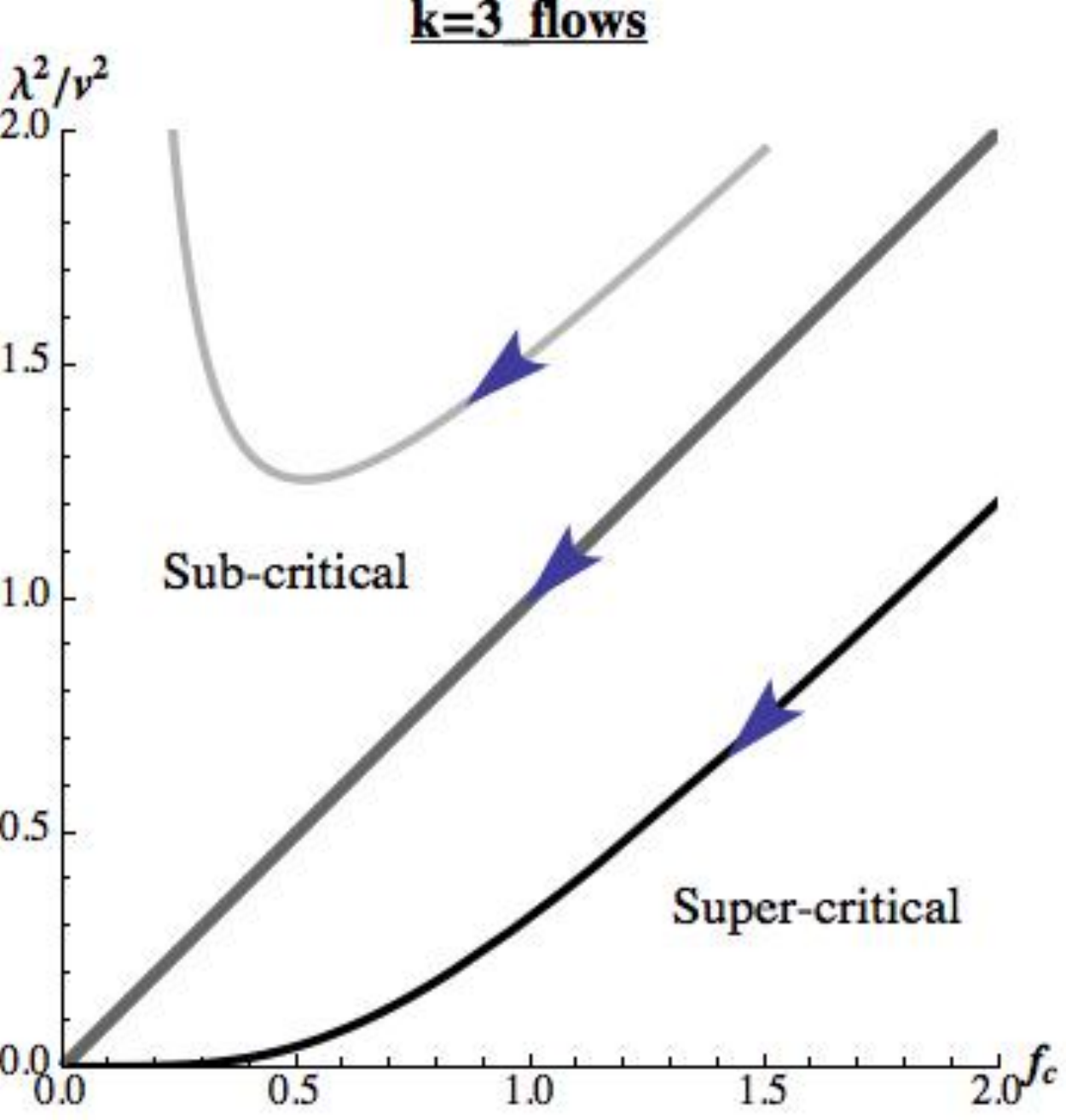} \hfil 
\includegraphics[width=0.328 \textwidth]{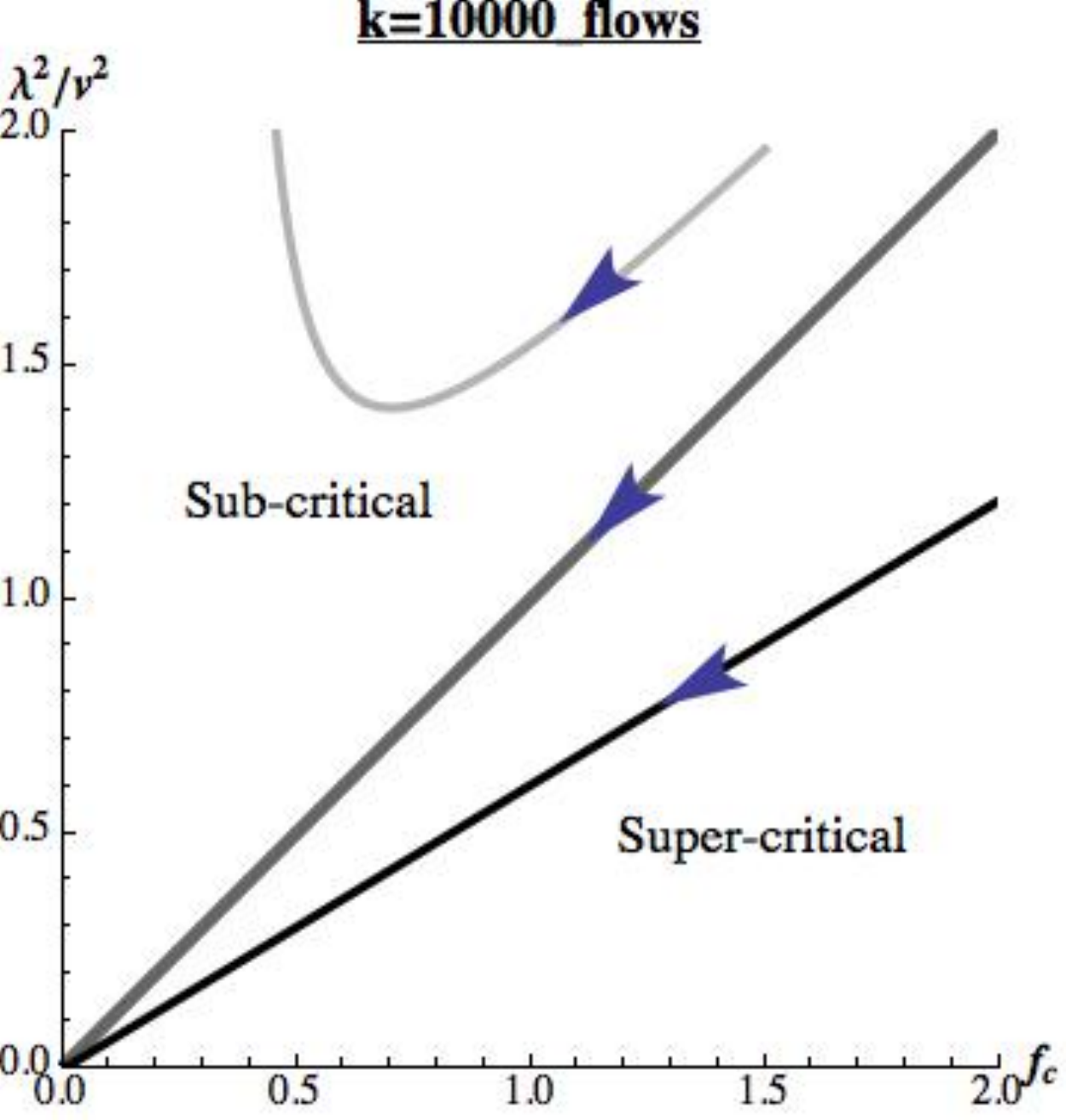} \\
({\bf A}) \hfil   \hspace{0.75in}  ({\bf B}) \hfil   \hspace{0.75in} ({\bf C}) \hfil 
\caption{Typical RG flows in three regions of parameter space. We plot $\bar \lambda^2/v^2$ against $f_c$ for (A) $k=2$, (B) $k=3$ and ({}C)  $k=10000$. In each plot, the lowest curve is a super-critical flow (with $f_c > \bar \lambda^2/v^2$), the middle curve is the critical flow (with $f_c = \bar \lambda^2/v^2$) and the top curve is a sub-critical flow (with $f_c < \bar \lambda^2/v^2$). We see that the super-critical flow is to finite $f_c$ for $k=2$ and to vanishing $f_c$ for $k >2$, and that $\bar \lambda^2/v^2$ diverges along the sub-critical flows, because $v \to 0$.}    
\label{fig:phases}
\end{figure}

We will now discuss the characteristic behaviours of the  RG flows for $k >1$.  Given the RG invariant condition \eqref{RGdiv}, $f_c = \bar \lambda^2/v^2$ divides the space of RG flows into three categories : subcritical where $f_c < \bar \lambda^2/v^2$, critical where $f_c = \bar \lambda^2/v^2$ and supercritical where $f_c > \bar \lambda^2/v^2$.\footnote{Note that subcritical includes the simplest analogue of the $k=1$ discussion of \cite{sitte}, $f_c=0$.}  This phase structure can clearly be seen in figure \ref{fig:phases}. 

The supercritical and critical RG flows, with $\bar \lambda^2 \leq v^2 f_c$, cannot reach $v=0$ at finite RG scale, since, as we discussed below \eqref{krg4},  $\bar \lambda \neq 0$ at finite scale if it was non-zero initially. However subcritical flows, with $\bar \lambda^2 > v^2 f_c$, could end at finite scale. Thus, since all parameters are decreasing along RG flows, we see that the supercritical and critical flows always extend to arbitrary $b$, with the parameters either going to zero or approaching finite values in the IR. Subcritical flows are harder to analyse, as the flow may terminate at some finite scale where $v$ vanishes, and so we cannot always use an IR expansion. 

\begin{figure}[t]
\begin{center}
\includegraphics[width=0.328 \textwidth]{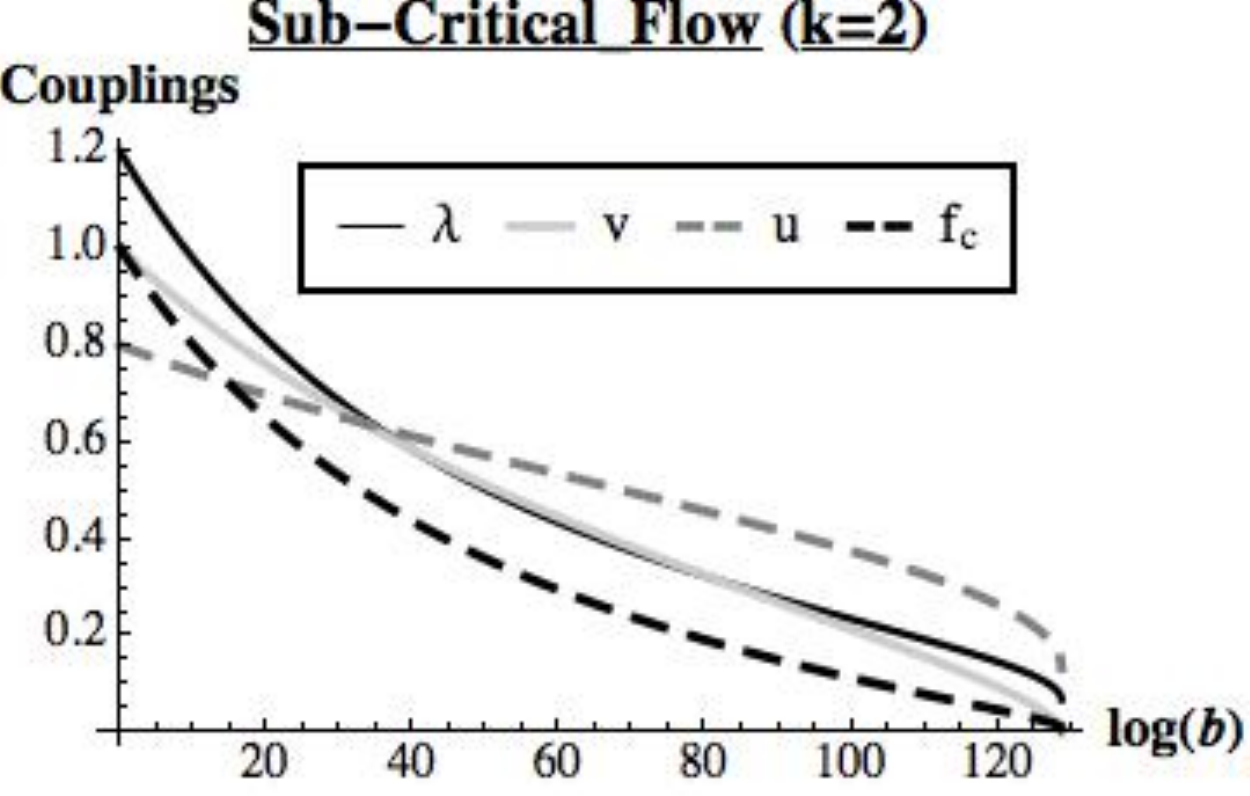} \hfil 
\includegraphics[width=0.328 \textwidth]{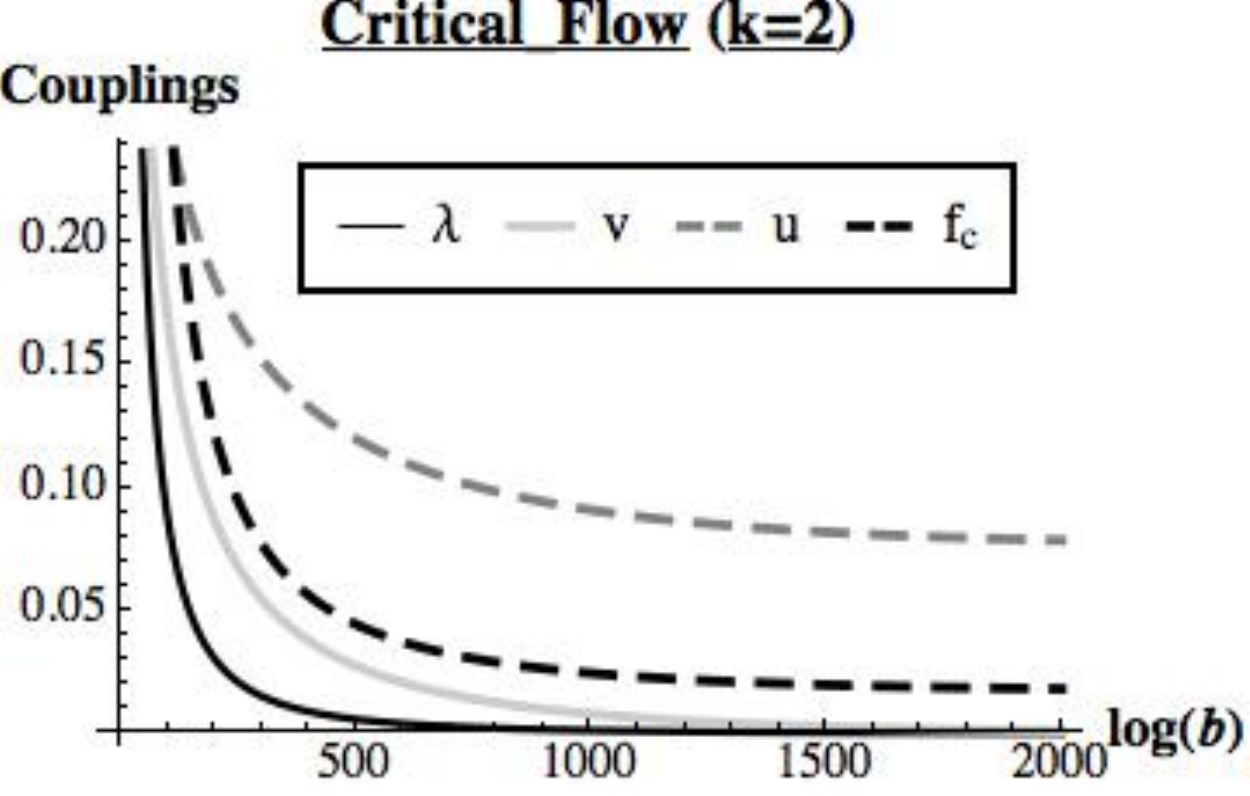} \hfil 
\includegraphics[width=0.328 \textwidth]{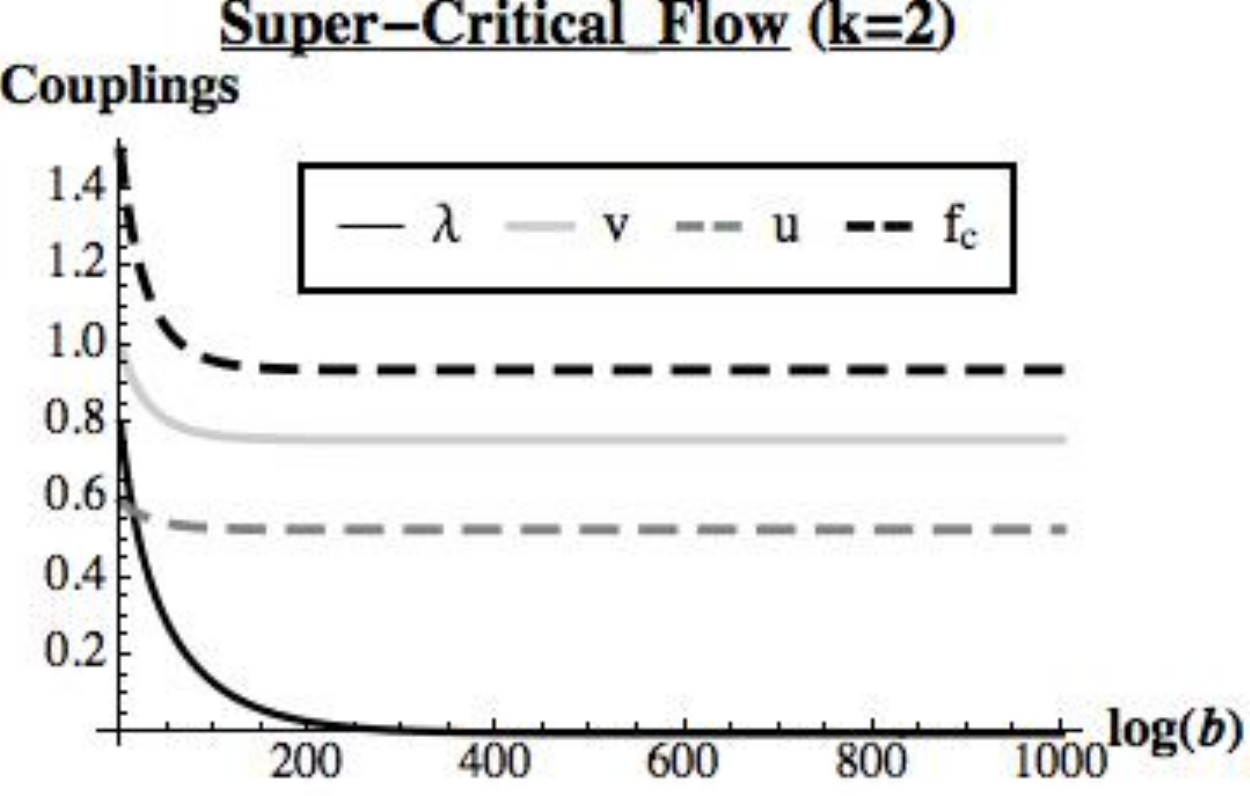} \\
({\bf A}) \hfil   \hspace{0.75in}  ({\bf B}) \hfil   \hspace{0.75in} ({\bf C}) \hfil 
\caption{RG flow for $k=2$, in the $z=1$ scheme. ({\bf A}) Sub-critical flow: $\lambda_0 = 1.2, \, v_0 =1, \, u_0 = 0.8, \, f_{c0} =1$. We see that $v \to 0$ at a finite scale. ({\bf B}) Critical flow: $v_0 =1, \,  u_0 = 0.6, \, f_{c0}= 4, \, \lambda_0 = \sqrt{fc_0 \, v_0^2}$. Here $\lambda, v \to 0$ in the IR, but $f_c, u$ remain finite. ({\bf C}) Super-critical flow: $\lambda_0 = 0.8, \, v_0 = 1, \, u_0 = 0.6, \, f_{c0} = 1.5$. Here $\lambda \to 0$ in the IR but the other couplings are finite.
}
\label{k=2Flows}
\end{center}
\end{figure}

\begin{figure}[t]
\begin{center}
\includegraphics[width=0.328 \textwidth]{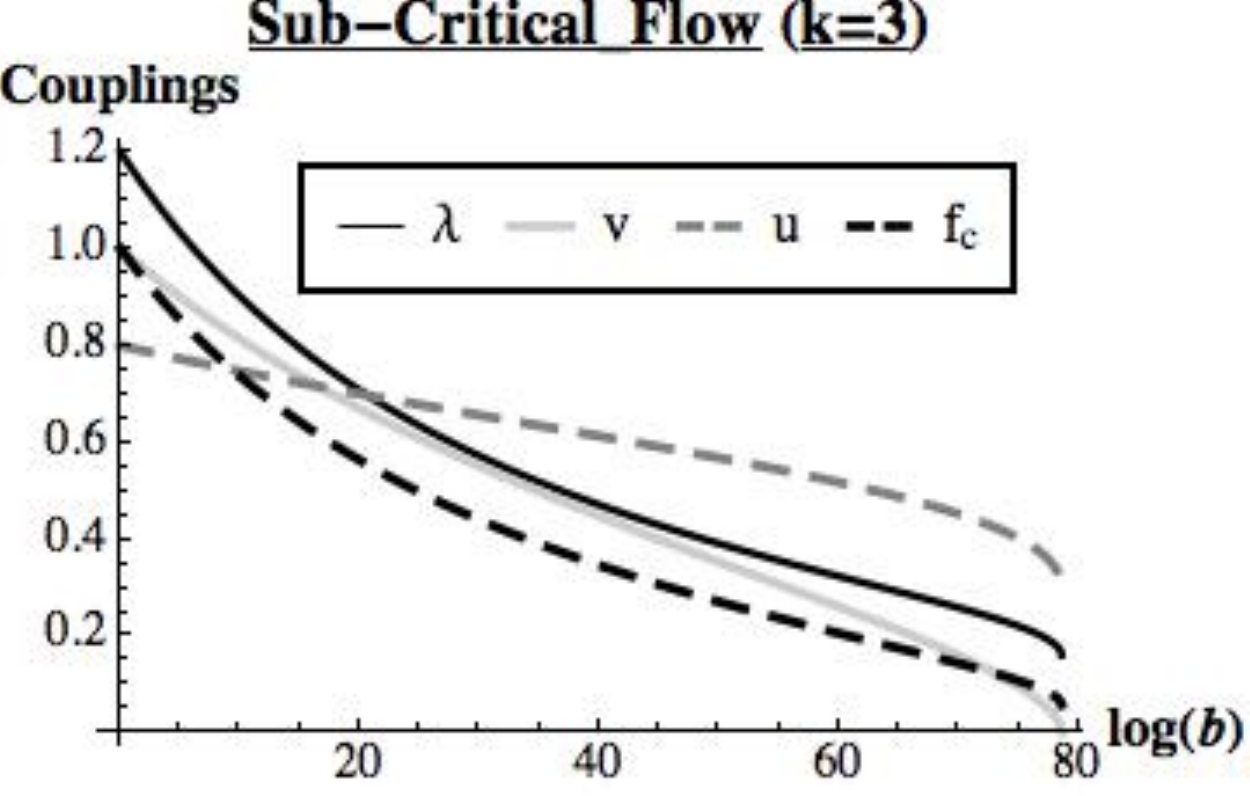} \hfil 
\includegraphics[width=0.328 \textwidth]{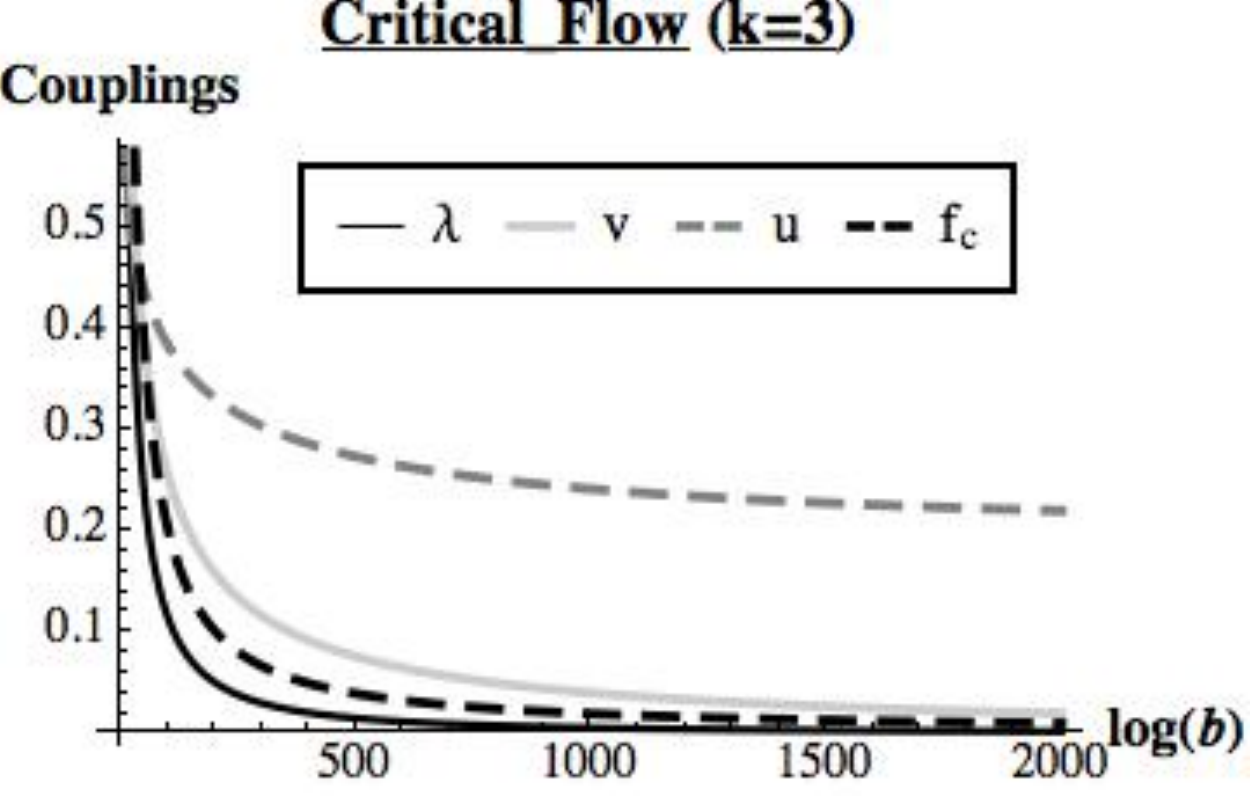} \hfil 
\includegraphics[width=0.328 \textwidth]{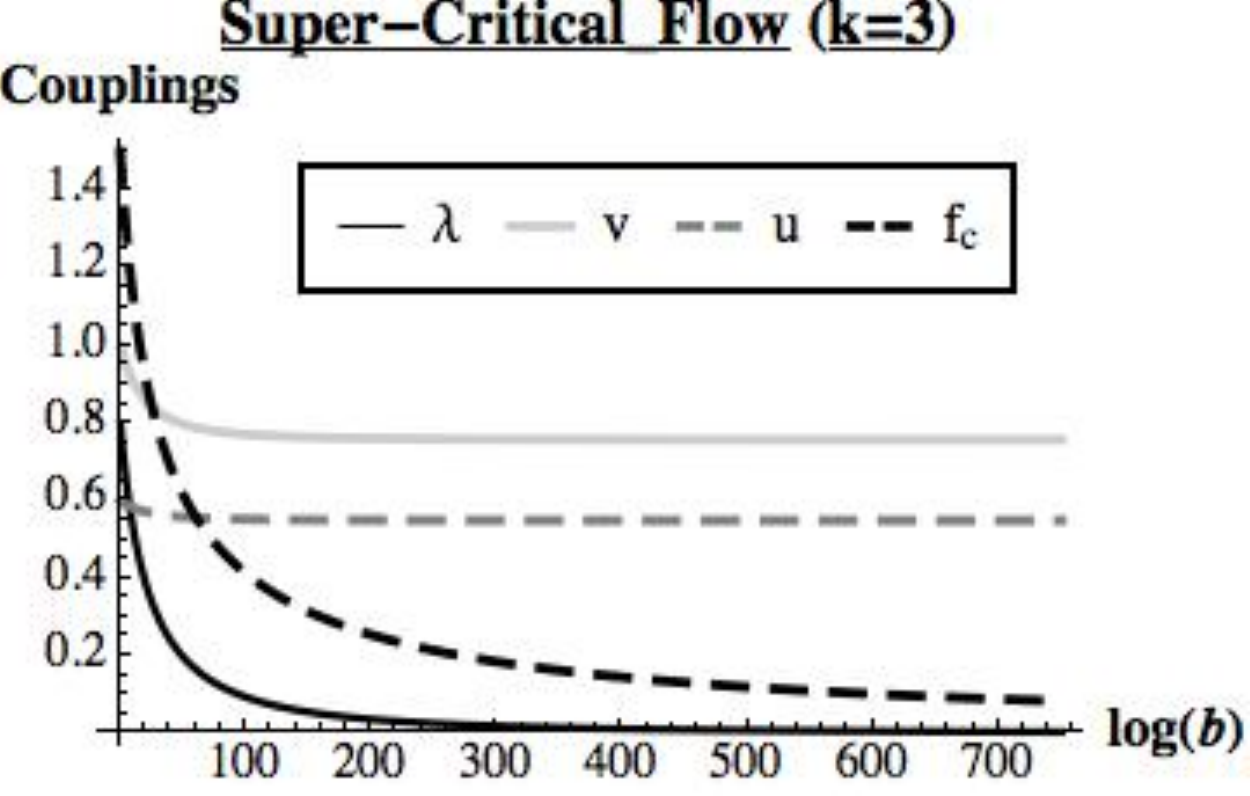} \\
({\bf A}) \hfil   \hspace{0.75in}  ({\bf B}) \hfil   \hspace{0.75in} ({\bf C}) \hfil 
\caption{RG flow for $k=3$ in the $z=1$ scheme. ({\bf A}) Sub-critical flow: $\lambda_0 = 1.2, \, v_0 = 1, \, u_0 = 0.8, \, f_{c0} = 1$. Here we see again that $v \to 0$ at a finite scale. ({\bf B}) Critical flow: $v_0 =1, \, u_0 = 0.6, \, f_{c0} = 2, \, \lambda_0 = \sqrt{f_{c0} \, v_0^2}$. Here $u$ remains finite as the other couplings flow to zero in the IR. The difference from $k=2$ is that now $f_c \to 0$ in the IR.  ({\bf C}) Super-critical flow: $\lambda_0 = 0.8, \, v_0 = 1, \, u_0 = 0.6, \, f_{c0} = 1.5$. Here $u, v$ are finite, but $\lambda, f_c \to 0$ in the IR.
}
\label{k=3Flows}
\end{center}
\end{figure}

\begin{figure}[t]
\begin{center}
\includegraphics[width=0.328 \textwidth]{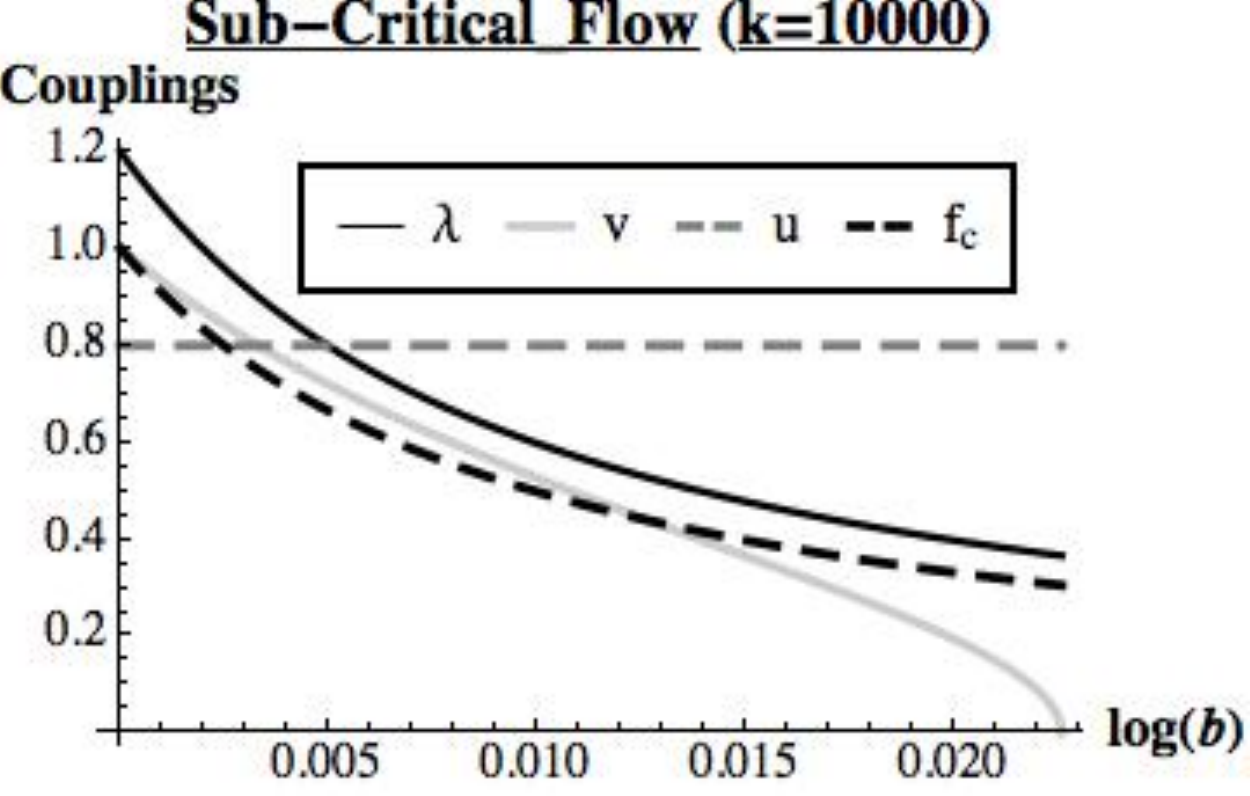} \hfil 
\includegraphics[width=0.328 \textwidth]{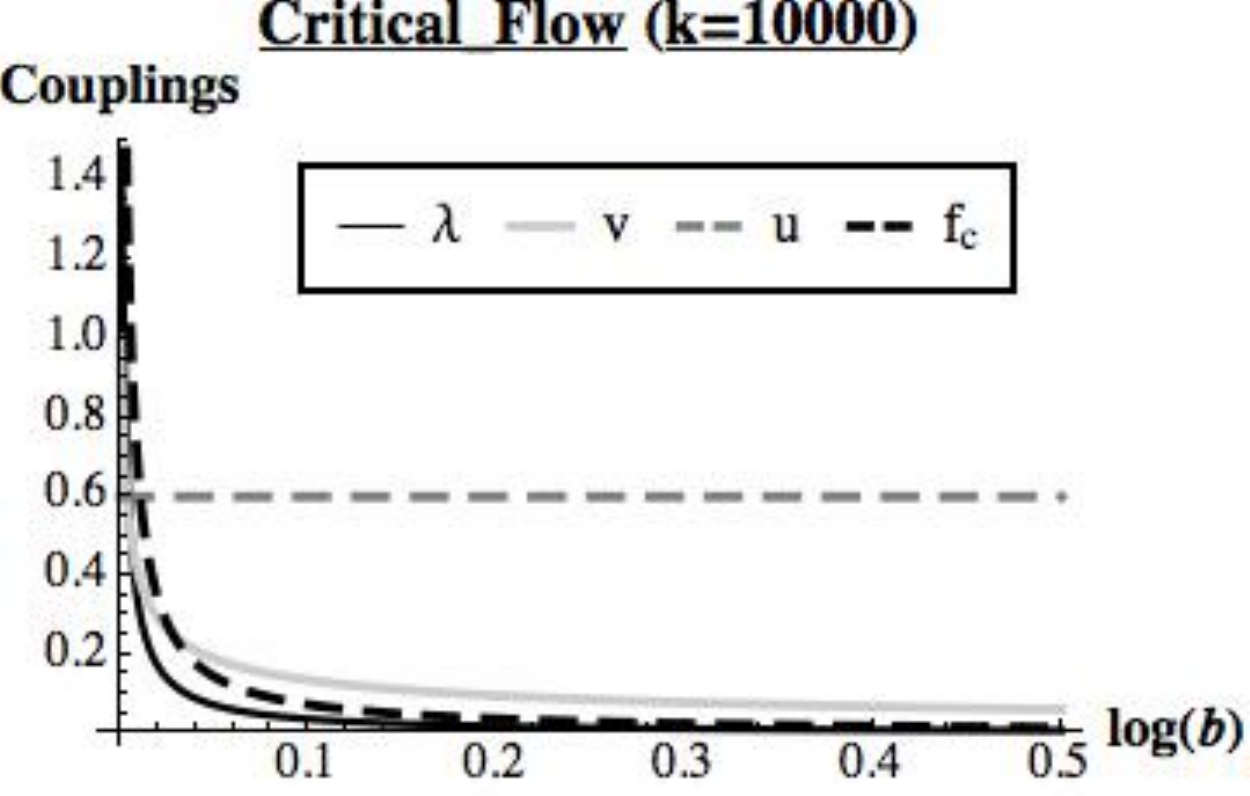} \hfil 
\includegraphics[width=0.328 \textwidth]{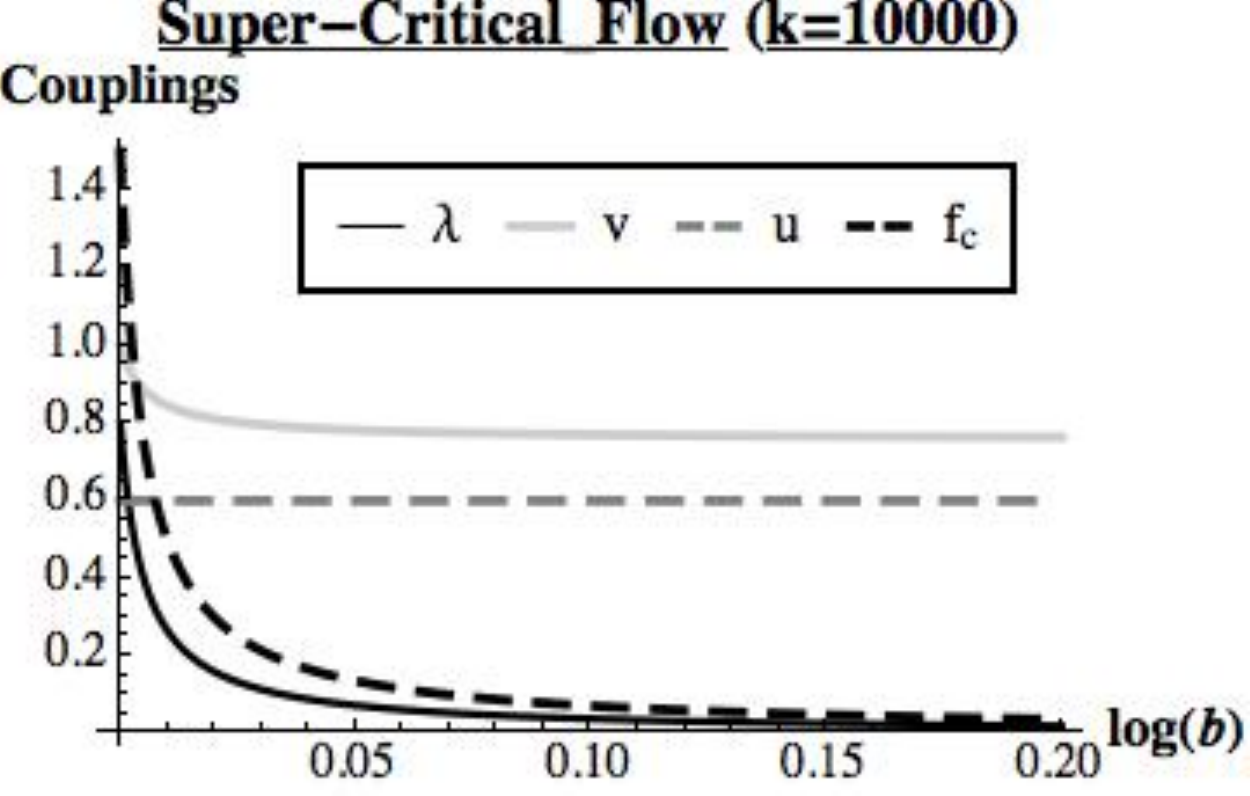} \\
({\bf A}) \hfil   \hspace{0.75in}  ({\bf B}) \hfil   \hspace{0.75in} ({\bf C}) \hfil 
\caption{RG flow for $k=10000$ in the $z=1$ scheme. We see that $u$ does not run to leading order in $k$ at large $k$. ({\bf A}) Sub-critical flow: $\lambda_0 = 1.2, \, v_0 = 1, \, u_0 = 0.8, \, f_{c0} = 1$. Here we clearly see $v \to 0$ at a finite scale with the other couplings remaining finite.  ({\bf B}) Critical flow: $v_0 =1, \, u_0 = 0.6, \, f_{c0} =4, \, \lambda_0 = \sqrt{f_{c0} v_0^2}$. All the parameters apart from $u$ run to zero. ({\bf C}) Super-critical flow: $\lambda_0 = 0.8, \, v_0 = 1, \,  u_0 = 0.6, \, f_{c0} = 1.5$. We see that $u$ and $v$ remain finite, while $\lambda$ and $f_c$ run to zero.}
\label{k=10000Flows}
\end{center}
\end{figure}

Since the RG equations are a set of first-order ODEs for the couplings, we can easily solve them numerically;  the results are presented in Figs. \ref{k=2Flows}, \ref{k=3Flows} and \ref{k=10000Flows} and summarised in Fig.~\ref{fig:phases}. We supplement this numerical analysis with analytical calculations for a few simple cases. 

\paragraph{Analytic calculations in a fixed $u$ scheme:} Setting $\bar \lambda^2 = f_c v^2$, the critical flows are two parameter flows in the fixed $u$ RG scheme. The relative flow $f_c(v)$ for critical flows can be determined for any $k$ by dividing \eqref{ukrg4} by \eqref{ukrg1}. This gives  
\begin{equation} 
\frac{\partial \log f_c}{\partial v} =   \frac{2(u+v)}{v} \frac{(k-2) (u+v) + 2v}{k (u+v)^2 - 4uv}. 
\end{equation}
Integrating this relation tells us that 
\begin{equation} 
\log f_c = \frac{2(k-2)}{k} \log v + \ldots, 
\end{equation}
where the dots stand for terms that remain finite when $v \to 0$. Thus, $f_c$ will vanish when $v$ vanishes, unless $k=2$. For $k=2$,
\begin{equation} 
\frac{\partial \log f_c}{\partial v} =  2 \frac{u+v}{u^2 + v^2} \quad \Longrightarrow \quad
\log f_c = \log (u^2 + v^2) + 2 \tan^{-1} \left( \frac{v}{u} \right) + C, 
\end{equation}
so that $f_c$ remains finite for all $v$. From \eqref{ukrg1}, the running of $v$ is then
\begin{equation} 
\frac{\partial v^2}{\partial \log b} = - {f_c v^2 \over \pi^3} \frac{(ku^2 + (2k-4) uv + kv^2)}{4u (u+v)^2}.
\end{equation}
For $k=2$, where $f_c\to f_c^{\text{IR}}$ is constant in the IR, the velocity goes to zero as a power law, $v \sim b^{-\alpha}$, with $\alpha = -f_c^{\text{IR}}/2\pi^3 u$. For $k >2$, we had $f_c \sim v^{2\beta}$, with $\beta = \frac{k-2}{k}$, so in the IR $v^{-2 \beta} \sim \log b$, and the velocity goes to zero logarithmically. It is interesting that $k=2$ seems to be a transitional case.   

 The absence of the last term in \eqref{ukrg4} for $k=2$ allows us to make further progress, even away from the critical flow. Dividing \eqref{ukrg4} by \eqref{ukrg1}, we can integrate the relative flow $f_c(v)$ for all $k=2$ flows, giving 
\begin{equation} 
\log f_c =  \log (u^2 + v^2) +2 \tan^{-1} \left( \frac{v}{u} \right) + C, 
\end{equation}
so that $f_c$ remains finite. For the supercritical case, we can assume the second term in \eqref{ukrg3} will dominate over the first one, so 
\begin{equation} 
\frac{\partial \bar \lambda}{\partial \log b} \approx  - \frac{f_c \bar \lambda}{4 \pi^3 u}.
\end{equation} 
Since $f_c\to f_c^{\text{IR}}$ is constant in the IR, $\bar \lambda$ will go to zero as a power law, $\bar \lambda \sim b^{-\alpha/2}$, where $\alpha$ is the same power as before. Plugging this into \eqref{ukrg1} will then give that $v$ will generically approach a constant, 
\begin{equation} 
v^2 \approx A + B b^{-\alpha}. 
\end{equation} 
The critical flow corresponds to the special case with $A=0$ where $v^2 \to 0$.

 Subcritical flows are harder to understand, because we don't know whether the first term in \eqref{ukrg3} is important. A special case which we can analyse explicitly is $f_c =0$. Dividing \eqref{ukrg3} by \eqref{ukrg1} gives
 \begin{equation} 
\frac{\partial \log \bar \lambda}{\partial v} = \frac{2(v - u)}{k(u^2 + v^2)+(2k-4)uv}, 
\end{equation} 
so the relative flow $\bar\lambda(v)$ is
 \begin{equation} 
\log \bar \lambda = - \frac{2}{k}\sqrt{k-1}\tan^{-1} \left(\frac{(k-2)u + kv}{2\sqrt{k-1}u}\right) + \frac{1}{k}\log\left(k(u+v)^2-4uv\right) + C.
\end{equation} 
Thus, $\bar \lambda$ remains finite for all $v$, including $v\to 0$. Plugging into \eqref{ukrg1}, we see that the RHS is bounded away from zero, so that $v^2$ will reach zero at a finite RG scale. This is qualitatively similar to the behaviour found in \cite{sitte} for $k=1$ when $u >v$; here we see that we get this behaviour for any velocities on the flow with $f_c =0$. In the numerical results in Fig.~\ref{k=2Flows}, we see that this behaviour is generic for the subcritical flows. 

\paragraph{Physical velocities and couplings:} As in \cite{sitte}, we can convert back from the fixed $u$ scheme we have used in our analytic discussion of the RG flows to the physical velocities (that is, to $z=1$) by multiplying by the factor
 \begin{equation} 
\Omega = \exp \left( \int^{\log b} (1-z(b')) d \log b' \right),
\end{equation} 
where $z$ is given in \eqref{uz}. Unlike in \cite{sitte}, this $\Omega$ is a finite factor for the flows we have analysed. For the critical flow, with $k>2$, 
 \begin{equation} 
1 -z \sim \frac{1}{(\log b)^{1+\frac{1}{\beta}}},
\end{equation} 
so the integral is finite. For $k=2$, the critical and supercritical flows have $1-z \sim e^{-\alpha \log b}$, so the integral is again finite. When $v$ goes to zero at a finite RG scale $b_c$ we have $v^2 \sim \log b_c - \log b$, so $1-z \sim (\log b_c - \log b)^{-1/2}$ and the integral again remains finite. As a consequence, the physical velocity $u$ remains finite along all these flows, and the IR asymptotics of the physical $v$ and the couplings is as described above.

\paragraph{Numerical calculations and large $k$ limit: } It is difficult to solve the RG equations analytically for $k>2$ non-critical flows, but straightforward to solve them numerically. Representative numerical plots for $k=3$ are given in Fig.~\ref{k=3Flows}; higher values are qualitatively similar. We see that the main difference from the $k=2$ case is that $f_c$ flows to zero in the IR.  In the limit of large $k$, the final terms in \eqref{krg1} and \eqref{krg3}-\eqref{krg4} will dominate over the other contributions, and $u$ will not run at leading order in $k$. It is then straightforward to solve the RG equations analytically. We postpone the detailed discussion to section~\ref{adscft}, as it is essentially equivalent to the solution obtained holographically (see equations \eqref{RGhol}). We present numerical results for a representative large $k$ case in Fig.~\ref{k=10000Flows}.

\paragraph{$k=2$ summary:}  Of the generalizations we have considered, perhaps the most interesting is $k=2$, which can be realized experimentally.    We have found that the flows in this case are qualitatively different from those with $k=1$ (Fig.~\ref{fig:phases}A,\ref{k=2Flows}).   Firstly, the IR fixed point for supercritical and critical flows is not generically relativistic, as $v$ runs relative to $u$. The physical velocity $u$ remains finite along all the $k=2$ flows, while the physical velocity $v$ remains finite for supercritical flows, goes to zero logarithmically in the IR for critical flows, or vanishes at a finite scale for subcritical flows. The coupling $f_c$ remains finite in all cases, while $\bar \lambda$ vanishes in the IR for supercritical or critical flows, remaining finite when $v$ vanishes in the subcritical case. 
 
 \subsection{The large $c$ model} \label{legCmdl}

The general RG equations for $k >2$ are rather difficult to analyse, but we can see that they simplify in the large $k$ limit, where the contributions from fermion loops will dominate over the other terms. This motivates us to consider a more general large central charge model:  we keep the scalar field $\phi$ describing the Luttinger liquid, but replace the $k$ fermions with a general 2d CFT of large central charge $c$, having a dimension one operator ${\cal O}$ which we couple to $\phi$ through an interaction of the same form as above. More explicitly, we have a scalar field  $\phi$ with the Lagrangian 
\begin{equation} \label{lutt}
\mathcal S_{LL} = \int d^2 x \, (\partial_t \phi^2 + v^2 \partial_x \phi^2),
\end{equation}
which provides the bosonised description of a Luttinger liquid.   We introduce a CFT sector with a dimension $(1/2,1/2)$ operator $\mathcal O$, which generalises the fermion bilinear $\psi^\dagger \psi$ in the previous discussion.  We add an interaction 
\begin{equation} \label{leq}
{\cal L}_{int} = -\tilde \lambda \, \partial_x \phi  \, \mathcal  O + F_c \, \mathcal O^2,
\end{equation}
including both a coupling to  the Luttinger liquid and the marginally irrelevant deformation $\mathcal O^2$.  We introduced new names for the interaction parameters because here and henceforth we adopt the standard CFT normalisation for $\mathcal O$ where the two-point function is $\langle \mathcal O(x) \mathcal O(0) \rangle = |x|^{-2}$. Then, if we specialise to the case where the CFT describes $k$ fermions, in which case $c = k/2$ and $\mathcal O = \frac{1}{\sqrt k} \mathcal O_M$, we learn that $\tilde \lambda = \sqrt{k} \bar \lambda$ and $F_c = k f_c$.  

The couplings whose RG runnings we are interested in understanding are $\tilde \lambda$, $F_c$, and the velocities of the two sectors. The velocity $v$ of the Luttinger liquid appears as a parameter in the Lagrangian \eqref{lutt}. There is a velocity $u$ characterising the CFT sector; we have implicitly set $u=1$ by choice of units at this stage, but it will run after we introduce the coupling \eqref{leq}.  We will discuss the holographic determination of its running later. 

We can work directly with this coupled theory, or given that $\phi$ is a free field with a Gaussian path integral, we can integrate it out explicitly  by solving its equation of motion.\footnote{This allows us to avoid the question of how the degrees of freedom of the Luttinger liquid would be explicitly realized in an AdS model.  This could have been alternatively treated in the semi-holographic approach of \cite{Faulkner:2010tq}.   See \cite{finnjoan,iqbalfaulkner} for explicit discussions of Luttinger liquids in the AdS/CFT context.} Doing so transforms the coupling \eqref{leq} into a momentum-dependent (and non-local) double-trace deformation $ F \mathcal O^2$, with coefficient  
\begin{equation} \label{Feq}
F = F_\lambda \frac{ v^2 k_x^2}{v^2 k_x^2 + k_t^2} + F_c,
\end{equation}
where $F_\lambda = -\tilde \lambda^2/v^2$. Due to the negative contribution of $F_\lambda$ to the CFT Hamiltonian, we
would expect this deformation to give a well-defined theory only if $F_c + F_\lambda \geq 0$ and to lead to a dynamical instability otherwise.

Indeed, as in the previous field theory discussion, we will find that the holographic RG flows are divided into two regions by the critical line $F_\lambda + F_c=0$. For $F_\lambda + F_c < 0$, the velocity $v$ vanishes at a finite RG scale - this is presumably associated with the expected instability mentioned above. Along the critical line $v$ and $F_c$ flow to zero in the IR, whereas for  $F_\lambda + F_c > 0$, $v$ flows to a finite value in the IR, and $F_c$ still flows to zero. The velocity $u$ of the CFT is not renormalised at leading order in $c$. 

\section{AdS/CFT approach}
\label{adscft}

Many strongly coupled CFTs with large central charge $c$ have a dual description in terms of gravity in asymptotically anti-de Sitter (AdS) space.   The AdS/CFT correspondence \cite{Maldacena:1997re} gives a procedure to compute the RG flow of the deformations of such theories in an expansion in inverse powers of $c$. We will briefly review this method below; excellent comprehensive reviews include  \cite{Hartnoll:2009sz,McGreevy:2009xe,Adams:2012th}.  We will apply it to study the flow of a 2d CFT deformed by the marginal double trace operator \eqref{Feq} that arises from integrating out a Luttinger liquid as above.   As we will see, the result is independent of the details of the CFT and hence we expect the qualitative aspects of the flow to be universal, at least for theories with sufficiently large central charge and coupling. 

The AdS/CFT correspondence provides a  technique for computing the partition function $Z_{\text{CFT}}[J] $ of deformations of a CFT   
\begin{equation}
\mathcal L_{CFT} \to \mathcal L_{CFT} + \sum_a J_a(x) \mathcal O_a (x),
\label{defCFT}
\end{equation}
where $J_a(x)$  are couplings (possibly position-dependent) and $\mathcal O_a$ are operators in the theory.    Correlation functions are obtained by differentiating this partition function with respect to $J_a$.       We are interested in analyzing how the couplings $J_a$ run with  scale.

The dictionary for carrying out this computation is as follows.  First, the vacuum of the theory corresponds to empty anti-de Sitter space.  For a two-dimensional CFT this is AdS${}_{3}$ space
\begin{equation} \label{ads}
  ds^2 = L^2 \frac{dz^2 + dt^2 + dx^2}{z^2}\, .
\end{equation}
The scaling symmetry of the CFT is realized as the isometry $x^\mu \to \lambda x^\mu$, $z \to \lambda z$. Position in the radial direction $z$ is thus associated with RG scale in the CFT; in particular, the asymptotic behaviour as $z \to 0$ is associated with local excitations in the CFT, and $z \to \infty$ corresponds to the deep infrared of the CFT. 

Next, in theories with an AdS$_3$ dual all operators can be generated as sums of products of  ``single-trace''  operators $\{ {\cal O}_i \}$, each of which corresponds to a field $\Phi_i$ that propagates on AdS space.  For scalars, the operator dimension $\Delta_+$ of ${\cal O}_i $ is related to the mass $m$ of $\Phi_i$ as 
\begin{equation}
  \Delta_\pm = 1 \pm  \sqrt{1 + 4m^2L^2}.
 \label{eq:massconf}
\end{equation}
($\Delta_-$ is a parameter that will become useful later.)   Our large central charge model involved an operator $\mathcal O$ with conformal dimension 1, so that $\mathcal O^2$ has dimension 2, making the couplings $F_c$ and $F$ in \eqref{leq} and \eqref{Feq} marginal. This corresponds to $m^2 = -1/L^2$.  This mass saturates the so-called  Breitenlohner-Freedman (BF) bound, and is the lowest mass in AdS space that leads to a stable scalar field \cite{Breitenlohner:1982jf}. 

The AdS/CFT correspondence states that the path integral in AdS space with particular boundary conditions for fields is related to the CFT partition function in the presence of sources \cite{Gubser:1998bc,Witten:1998qj}. Below we will describe  how the boundary conditions are related to the sources.  We take our AdS action for the metric and $\Phi$, the AdS scalar field dual to the operator ${\cal O}$, to be
\begin{equation} \label{act}
S_{\text{cl}} = \frac{1}{16\pi G_N} \int d^{3} x \sqrt{-g} (R + \frac{2}{L^2}) + \frac{1}{2}\int d^{3} x \sqrt{-g} \, [ (\nabla \Phi)^2 + m^2 \Phi^2] \, ,
\end{equation} 
where $G_N$ is Newton's constant. The AdS length $L$ appearing in the cosmological constant term is related to the central charge in the CFT via
\begin{equation}
c=   \frac{3 L}{2G_N}  \gg 1\, .
\label{Landc}
\end{equation}
The large $c$ limit is thus a classical limit for the path integral with action \eqref{act}. Henceforth, we will choose units such that $L=1$; so the central charge $c \sim 1/G_N$. 

In this classical limit, the CFT partition function $Z_{\text{CFT}}[J]$ is approximated by calculating the action \eqref{act} on a classical solution. For a scalar field,  the classical equation of motion is 
\begin{equation} \label{eom}
z^3 \partial_z (z^{-1} \partial_z \Phi) + z^2 \partial_t^2 \Phi + z^2 \partial_x^2 \Phi - m^2 \Phi = 0.
\end{equation}
The solution for a scalar saturating the BF bound takes the asymptotic form
\begin{equation} \label{bfs}
\Phi \sim \Phi_+ \, z + \Phi_-  \, z \log (\Lambda z)
\end{equation}
as $z \to 0$. For $m^2$ exceeding the BF bound,  $\Phi \sim \Phi_- z^{\Delta_-} + \Phi_+ z^{\Delta_+}$, where $\Delta_\pm$ are given in (\ref{eq:massconf}). A simple boundary condition is $\Phi_- = J$, and
\begin{equation}
 Z_{\text{CFT}}[J] \approx e^{-S_{\text{cl}}[J]}.
\label{eq:classical}
\end{equation}

For fixed $\Phi_-$ we determine a solution $\Phi$ of the bulk equation of motion by imposing regularity in the interior of the spacetime. In Euclidean signature the condition of regularity in the bulk fixes $\Phi_+$ uniquely.  The subleading mode $\Phi_+$ can be identified with the expectation of the operator ${\cal O}$ (of conformal dimension $\Delta_+$) that develops in response to a perturbation of the theory around the conformal vacuum by a coupling $J(x) {\cal O}(x)$.  


The above discussion was extended to understanding the bulk description of ``double-trace'' operators like  $\mathcal O^2$ in \cite{Aharony:2001pa,Witten:2001ua,Berkooz:2002ug}.  In particular, adding a double-trace deformation corresponds to a boundary condition  $\Phi_-=F \Phi_+$. To be more careful, we should write the boundary conditions on a cutoff surface of fixed $z_0$; this then corresponds to 
\begin{equation}
\label{eq:bc}
z_0^{-\Delta_-} \Phi(z_0) = F(z_0) \, z_0^{-\Delta_+} \frac{ \Delta_- \Phi(z_0) - z \partial_z \Phi(z_0)}{\Delta_- - \Delta_+}.  
\end{equation}
For general values of $m^2$, the scaling transformation $x^\mu \to \lambda^{-1} x^\mu$, $z \to \lambda^{-1} z$ scales $\Phi_+ \to \lambda^{\Delta_+} \Phi_+$ and $\Phi_- \to \lambda^{\Delta_-} \Phi_-$, so $F \to \lambda^{d - 2 \Delta_+} F$, as we would expect for the coupling to an operator $\mathcal O^2$ of dimension $2 \Delta_+$.

In our case, where the operator saturates the BF bound, the scaling is a little more complicated; the asymptotic form \eqref{bfs} implies that the scaling is
\begin{equation}
\Phi_+ \to \lambda^{-1} ( \Phi_+ + \Phi_- \log \lambda), \quad \Phi_- \to \lambda^{-1} \Phi_-,
\end{equation}
so the RG flow of $F = \Phi_- / \Phi_+$ is  \cite{Witten:2001ua}
\begin{equation} \label{wflow}
F(\mu)= \frac{F(\Lambda)}{1+F(\Lambda) \log(\Lambda/\mu)},
\end{equation}
where we have re-expressed the scaling in terms of a ratio of energy scales $\Lambda, \mu$. (Another useful reference on boundary conditions for scalars saturating the BF bound is \cite{Amsel:2011km}).

When $F(\Lambda) >0$, this says that the theory returns to the undeformed $F=0$ theory in the IR ($\mu \to 0$), i.e. ${\cal O}^2$ is marginally irrelevant. If we try to extend the RG running to $\mu > \Lambda$ we will encounter a Landau pole at some critical energy $\mu_c$ where $F \to \infty$.   Thus the theory is not defined at arbitrarily high scales.  
One should therefore think of this as describing just the running below some cutoff scale $\Lambda$.  
On the other hand, if $F(\Lambda) < 0$, the Landau pole occurs at $\mu_c < \Lambda$, so the running to the IR will encounter this pole. We interpret this as a signal of the instability of the theory with $F <0$. 

In \cite{Witten:2001ua} $F$ was implicitly assumed to be momentum-independent, but the analysis there does not depend in any way on this assumption.   So we can extract the running of the  couplings $F_c$, $v^2$ and $F_\lambda$ in our large $c$ model by simply plugging the expression \eqref{Feq} into \eqref{wflow}. This gives 
\begin{eqnarray} 
  F_c(\mu)&=& \frac{F_c(\Lambda)}{1+F_c(\Lambda)\log(\Lambda/\mu)}, \label{eq:fc} \\
  v^2(\mu) &=& v^2(\Lambda) \frac{1+(F_c(\Lambda)+F_\lambda(\Lambda))\log(\Lambda/\mu)}{1+F_c(\Lambda) \log(\Lambda/\mu)} ,\\
  F_\lambda(\mu) &=& \frac{1}{1+F_c(\Lambda)\,\log(\Lambda/\mu)}\biggl(\frac{F_\lambda(\Lambda)}{1+ \left(F_c(\Lambda)+F_\lambda(\Lambda)\right)\log(\Lambda/\mu)}\biggr). \label{FlRG}
\end{eqnarray}
In particular, notice
\begin{equation}
  F_\lambda(\mu) + F_c(\mu) = \frac{F_\lambda(\Lambda) + F_c(\Lambda)}{1 + \left(F_\lambda(\Lambda) + F_c(\Lambda)\right)\log(\Lambda/\mu)}.
\end{equation}
We see that the IR behaviour depends  on the sign of $F_\lambda + F_c$. The parameter space is divided into different regions by a critical line at $F_c + F_\lambda =0$.\footnote{In this approach we are reading off the beta functions of the field theory in terms of the the radial running of the boundary condition for an AdS bulk field.     A more Wilsonian approach, integrating out shells in spacetime to derive an effective IR theory, was followed in \cite{Heemskerk:2010hk,Faulkner:2010jy}.  (See \cite{pervijay} for a related earlier approach.)   The running couplings derived in \cite{Heemskerk:2010hk,Faulkner:2010jy} do not initially look anything like the running value of bulk solutions.   However, it is possible to choose an RG scheme in which the Wilsonian couplings are indeed directly related to the running values of bulk fields \cite{monicaalbion}.}  To compare to the previous field theory analysis, we can rewrite these flows in terms of beta functions, and trade the coupling $F_\lambda$ for $\tilde \lambda^2 = - F_\lambda v^2$. The beta functions are 
\begin{equation}
\begin{split}
\frac{\partial F_c}{\partial \log(\Lambda/\mu)} &= -F_c^2, \\
\frac{\partial v^2}{\partial \log(\Lambda/\mu)} &= - \tilde \lambda^2 ,\\
\frac{\partial \tilde \lambda}{\partial \log(\Lambda/\mu)} &= -F_c \tilde \lambda. 
\end{split}
\label{RGhol}
\end{equation}

Comparing these to the large $k$ limit of the field theory RG equations in \eqref{krg1}-\eqref{krg4}, we see that there is also no running of $u$ at leading order in $k$ in the field theory calculation, and that the functional form of the RG running of the other couplings is the same, apart from a factor of $k/4\pi^3u$ in the RHS of \eqref{krg1}-\eqref{krg4}. The factor of $k$ is due to the redefinitions $\tilde \lambda^2 = k \bar \lambda^2$, $F_c = k f_c$, while we have set $u=1$ in the present discussion.

Thus, the RG running of the coupling in the field theory will also be given, up to normalisation, by the solution \eqref{FlRG} read off from the holographic analysis. This agrees well with the behaviour obtained from the numerical analysis in  Fig.~\ref{fig:phases}C and Fig.~\ref{k=10000Flows}. In summary,
\begin{itemize}
\item For subcritical flows, $F_c + F_\lambda <0$, the theory reaches a pole at a finite scale $\mu_c$ where $1 + (F_c + F_\lambda) \log (\Lambda/\mu_c) = 0$. At $\mu_c$ the velocity $v$ vanishes, while $\tilde \lambda$ and $F_c$ remain finite (the divergence in $F_\lambda$ is just due to the vanishing of the velocity $v$). This pole is presumably associated with the expected instability of this theory. This includes in particular the case $F_c = 0$ which is the most straightforward generalisation of the one-fermion system studied in \cite{sitte}. Note that for $F_c=0$, the RG at leading order in $c$ affects only the velocity $v^2$. This agrees with our analysis of the $F_c=0$ flow at $k=2$, and this behaviour of the subcritical flows appears to be robust for $k >1$.  
\item Along the critical line $F_c + F_\lambda =0$, the velocity $v$ flows to zero in the IR, and the couplings $F_c$, $\tilde \lambda$ will also flow to zero. This agrees qualitatively with the $k=3$ flow in Fig.~\ref{k=3Flows}, so this behaviour appears robust for $k >2$. 
\item For supercritical flows with $F_c + F_\lambda >0$, the velocity $v$ flows to a finite value in the IR, while the couplings $F_c$, $\tilde \lambda$ still flow to zero.   This again agrees with the behavior seen in 
Fig.~\ref{k=3Flows}.
 \end{itemize} 
Thus, the holographic analysis has produced an RG behaviour which agrees with the large $k$ limit of the field theory and allowed us to extract the solution of the RG equations simply from the AdS behaviour. The behaviour is also consistent with field theory results at finite $k >2$. 

The AdS analysis does much more, however, in two respects. The first is that it computes the behavior for generic large $c$ CFT, under the only assumption of  a dimension 1 operator which couples to the full current, i.e. it is very robust to changes in the model\footnote{Even if the CFT is interacting we still expect to see a dimension 1 operator. Recall that we obtain the CFT by coupling a massive theory to the low band, then redefining the low band Luttinger liquid to be that of total electric charge by separating off the charge from the massive theory, and then going to the IR in the stripped off theory. However, the dimension 1 charge density operator is still there and couples to $\partial_x \phi$. Since the stripped-off CFT no longer has a $\U(1)$ symmetry, that operator is a $(1/2,1/2)$ operator rather than a current.}. The second is that goes beyond the one loop field theory, and sums up all the planar diagrams.  

What about  the running of the velocity $u$?  As noted in the field theory discussion, at large central charge $c$, $u$ does not run to leading order in $1/c$.    To explore the renormalisation of $u$, we therefore turn in the next section to the subleading corrections that are produced by back-reaction of the scalar field onto the bulk metric.

\section{Back-reaction and flow of $u$}
\label{br}

To explore the renormalisation of the CFT velocity $u$, we need to consider the back-reaction on the metric of the boundary condition for the scalar $\Phi$.  As we will see, this will first appear at subleading order in $1/c$. We calculate this back-reaction first for the momentum-independent case where we only consider $F_c$, and then in the general case with both $F_c$ and $F_\lambda$ turned on in the boundary condition. The calculation is similar to a previous AdS/CFT calculation of the back-reaction of a double-trace deformation \cite{Gubser:2002zh}, but differs in that we consider a scalar saturating the Breitenlohner-Freedman bound and  consider a momentum-dependent boundary condition. We find that the velocity $u$ decreases in the IR, but remains finite, as in the field theory analysis for $k >1$. 

\subsection{General formalism}

 We start with  a general ansatz for the spacetime metric
\begin{equation} \label{linmet}
  ds^2 = \frac{e^{2 \nu(z)} dt^2 + dx^2 + g(z) dz^2}{z^2} \,.
\end{equation}
Recall that $z$ is identified with scale in the CFT.   In view of this, we can  interpret the metric on surfaces of constant $z$ as defining the background geometry seen by the CFT as a function of scale. In this interpretation, the ratio of the $g_{tt}$ and $g_{xx}$ components of the metric, $e^{\nu(z)}$, is  the running value of the velocity $u$.  At leading order in $c$, the bulk metric is simply AdS$_3$, even if the double-trace deformation modifies the boundary condition for the scalar.   To see changes in the geometry we need to take into account the back-reaction of the quantum stress tensor of the scalar field on the metric, through Einstein's equations 
\begin{equation}
R_{\mu\nu} - \frac{1}{2} R g_{\mu\nu} + \frac{1}{L^2} g_{\mu\nu} = 16 \pi G_N \langle T_{\mu\nu} \rangle.
\end{equation}
 Because $G_N \sim 1/c$, we see that the back-reaction of the metric to the matter is a subleading effect in the $1/c$ expansion, consistent with what we saw in the large $c$ field theory calculation. 

The expectation value of the stress tensor is 
\begin{equation}
\langle T_{\mu\nu} \rangle  = \langle \partial_\mu \Phi \partial_\nu \Phi \rangle - \frac{1}{2} g_{\mu\nu} \langle (\partial \Phi)^2 \rangle - m^2 g_{\mu\nu} \langle \Phi^2 \rangle. 
\end{equation}
To compute this, we evaluate the bulk two-point function $\langle \Phi(x,z) \Phi(y,\hat z) \rangle$ with the deformed boundary condition, take the relevant derivatives, and then take the limit $x \to y, z \to \hat z$. There is a divergence in the coincident point limit, which we can remove by considering the difference between $\langle T_{\mu\nu} \rangle$ computed with the boundary condition determined by $F$, and $\langle T_{\mu\nu} \rangle_0$ computed with the boundary condition $F=0$, that is $\Phi_- = 0$. 

The scalar two-point function is a  solution of  the scalar equation of motion \eqref{eom} which is regular as $z \to \infty$ and satisfies the asymptotic boundary condition $\Phi_- = F \Phi_+$ as $z \to 0$, and which has a delta-function source at some $x =y, z = \hat z$. We will solve this equation by working in momentum space in the $t,x$ directions, with a delta-function source at $z = \hat z$. We write this solution, at a given momentum 2-vector $k$, $\Phi(k,z;\hat z)$ as 
\begin{equation}
\Phi = \left\{ \begin{array}{c} A_1 \Phi_1 \quad (z < \hat z) \\ B_2 \Phi_2 \quad (z > \hat z) \end{array} \right.,
\end{equation}
The solution of \eqref{eom} regular at the horizon $z \to \infty$ in AdS is simply $\Phi_2 = z K_0(k z)$, where we also use $k$ as a short hand notation for $|k|$. The solution satisfying the boundary condition at $z \to 0$ can be written as 
\begin{equation}
\label{eq:fsol}
\Phi_1 = z (a I_0(kz) + K_0(kz))\,,
\end{equation}
where $a$ is some arbitrary constant at this stage. We impose the boundary condition on some cutoff surface at a finite $z_0$. We think of $1/z_0$ as a UV cutoff for the field theory, so the momenta in the boundary directions will naturally satisfy $k z_0 \ll 1$, so that we can use the asymptotic form of this solution in imposing the boundary condition at $z_0$.\footnote{For points $\hat z$ far from the boundary, the dominant contribution to the one-loop vacuum diagram will come from such momenta.} Using the small argument expansion of the Bessel functions, for $z \approx z_0$, 
\begin{equation} \label{pasymp}
\Phi_1(z) \approx z \left(a - \gamma - \log\left(\frac{kz}{2}\right) \right).
\end{equation}
Imposing the boundary condition then gives 
\begin{equation} \label{a}
a=\log(kz_0)+A-1/F(z_0) ,
\end{equation}
where $A\equiv \gamma - \log 2$. Note that if we change the cutoff surface on which we impose the boundary condition, and calculate the value $F(z_1)$ on the new surface using the RG flow equation \eqref{wflow}, we obtain the same value for $a$, as we should: 
\begin{equation}
a=\log(kz_1)+A-1/F(z_1)=\log(kz_1)+A-{1+F(z_0)\log(z_1/z_0) \over F(z_0)}=\log(kz_0)+A-1/F(z_0). 
\end{equation}

The coefficients $A_1$, $B_2$ are then determined by requiring that $\Phi$ is continuous at $\hat z$ and its derivative has an appropriate discontinuity, so that $\partial_z^2 \Phi = z \delta(z-\hat z)$ as in \cite{Gubser:2002zh}. This gives
\begin{eqnarray}
   A_1\Phi_1(\hat z) - B_2\Phi_2(\hat z) &=& 0 \\
   A_1\partial_z\Phi_1(\hat z) - B_2\partial_z \Phi_2(\hat z) &=& \hat z,
 \end{eqnarray}
whose solution is 
 \begin{equation}
   A_1 = -\hat z\frac{\Phi_2(\hat z)}{W(\Phi_1, \Phi_2)}\,, \quad \quad B_2 = -\hat z \frac{\Phi_1(\hat z)}{W(\Phi_1, \Phi_2)}\,.
\end{equation}
The Wronskian $W(\Phi_1,\Phi_2) = \Phi_1 \partial_z \Phi_2 - \Phi_2 \partial_z \Phi_1$ of the solutions $\Phi_1$, $\Phi_2$ obtained above is $W(\Phi_1, \Phi_2) = - a \hat z$. Thus the Green's function is 
\begin{equation}
\Phi = \left\{ \begin{array}{c} \frac{1}{a} \Phi_2(\hat z) \Phi_1(z) \quad (z < \hat z) \\ \frac{1}{a} \Phi_1(\hat z)  \Phi_2(z)  \quad (z > \hat z) \end{array} \right. . 
\end{equation}

This gives us the Green's function in momentum space. To calculate the one-loop contribution to the stress tensor, we need to evaluate the Green's function in position space. This is given by an integral over momenta, 
\begin{equation}
\langle \Phi(x,z) \Phi(y, \hat z) \rangle = \int \frac{d^2 k}{4 \pi^2} e^{i \vec{k} \cdot (\vec{x}-\vec{y})} \frac{ z \hat z K_0(k \hat z) (a I_0(k z) + K_0(k z))}{a} \quad \quad z < \hat z,
\end{equation}
and a similar expression for $z> \hat z$. The Green's function will have a divergence at coincident points, but this is a UV effect (in spacetime) which is not sensitive to the modification of the boundary conditions. We cancel this divergence by considering the difference between the Green's function and the one for the undeformed theory with $F = 0$ boundary conditions (that is, the standard Dirichlet theory in AdS space). Since $F=0$ corresponds to $1/a = 0$, this just cancels the piece proportional to $I_0$, giving  
\begin{equation}
\label{eq:green}
\langle \Phi(x,z) \Phi(y, \hat z) \rangle_{ren}= \int \frac{d^2 k}{4 \pi^2} e^{i \vec{k} \cdot (\vec{x}-\vec{y})} \frac{ z \hat z K_0(k \hat z)\,K_0(k z))}{a} \quad \quad z < \hat z,
\end{equation}
which has a finite limit as $x \to y$, $z \to \hat z$. Note that in the integral over momenta, the exponential decay of the Bessel functions $K_0(kz)$ implies that the contribution comes mostly from momenta such that $kz \leq 1$, so if we consider a point far from the cutoff surface, so $z \gg z_0$, then $k z_0 \ll 1$, and our use of the asymptotic form in \eqref{pasymp} is justified. 

\subsection{Warm-up: momentum-independent boundary conditions}
\label{sec:wup}

As a warm-up, consider the calculation for a constant double-trace deformation (that is, just considering the coupling $F_c$, with $F_\lambda =0$). This is Lorentz-invariant, so its analysis is technically simpler than the momentum-dependent case.  It is also interesting for its relation to previous AdS/CFT calculations. Readers uninterested in the technical details who just want to see the results relevant to the zig-zag transition may want to skip this section on a first reading. 

Our aim in this section is to explore the difference between the present case, where the bulk scalar saturated the BF bound, and the discussion of \cite{Gubser:2002zh}, where the one-loop back-reaction due to a modified boundary condition for a scalar field with mass strictly above the Breitenlohner-Freedman bound was calculated.\footnote{In \cite{Gubser:2002zh} it was suggested that there would be no back-reaction as a result of turning on such a source, and indeed the scalar saturating the Breitenlohner-Freedman bound was used as a reference to calculate the behaviour above the bound. We will see that there actually is a flow, but as it involves a logarithmic running, the effect on  the computation in \cite{Gubser:2002zh} is negligible compared to the power-law running considered there.} The corrected bulk metric determines the running of the central charge in the field theory. We find that the central charge logarithmically approaches the value in the IR CFT. That is, the running is qualitatively similar to that in \cite{Gubser:2002zh}, but the behaviour is logarithmic rather than power law. The holographic running of the central charge was studied further in \cite{Gubser:2002vv,Hartman:2006dy}.

The components of the stress tensor we will be interested in are 
\begin{eqnarray}
\langle T_{tt} \rangle + \langle T_{xx} \rangle &=& - \langle\partial_z \Phi \partial_z \Phi\rangle + \frac{m^2}{z^2} \langle\Phi^2\rangle,  \nonumber \\
\langle T_{zz}\rangle &=& \frac{1}{2} ( \langle\partial_z \Phi \partial_z \Phi\rangle - \langle\partial_t \Phi \partial_t \Phi\rangle - \langle\partial_x \Phi \partial_x \Phi\rangle + \frac{m^2}{z^2} \langle\Phi^2\rangle), \nonumber
\end{eqnarray}
The expectation value $\langle\Phi^2\rangle$ is just the coincidence limit of \eqref{eq:green}; the derivative terms can be calculated by taking the derivatives of the renormalised Green's function \eqref{eq:green} and then taking the coincident point limit. We obtain
\begin{equation}
  \langle\partial_t \Phi\partial_t \Phi\rangle + \langle\partial_x \Phi\partial_x \Phi\rangle = \int \frac{d^2k}{4\pi^2 a} \left(k z\,K_0(kz)\right)^2, 
 \end{equation}
 \begin{equation}
  \langle\partial_z\Phi\partial_z\Phi\rangle = \int \frac{d^2k}{4\pi^2 a} \left(K_0(kz) + kz K_0'(kz)\right)^2.
  \end{equation}

Thus, the stress tensor components are 
\begin{equation}
\langle T_{\mu\nu} \rangle = \int \frac{k dk}{4 \pi a} H(kz),
\end{equation}
where for $\langle T_{tt} \rangle + \langle T_{xx} \rangle$,
\begin{equation} 
H(kz) = 2 \,  [ -(K_0(kz) + kz K_0'(kz))^2 + m^2 K_0(kz)^2],
\end{equation}
and for $\langle T_{zz} \rangle$, 
\begin{equation} 
H(kz) =  [ (K_0(kz) + kz K_0'(kz))^2 - k^2 z^2 K_0(kz)^2 + m^2 K_0(kz)^2].
\end{equation}
Recall that $a$ in \eqref{a} is also $k$-dependent. It is convenient to eliminate the apparent $z_0$ dependence in \eqref{a} by re-expressing $a$ in terms of the coupling $F$ at scale $z$, 
\begin{equation}
a = A + \log(kz) - \frac{1}{F(z)}. 
\end{equation}
The factor of $H(kz)$ is an exponentially decaying function of its argument, so the integral will be dominated by $kz \sim 1$. The RG flow \eqref{wflow} approaches $F \to 0$ in the IR, so if we assume that the scale $z$ is sufficiently far in the IR, then we can approximate $a \approx -1/F(z)$, and we can then explicitly evaluate the momentum integrals:
\begin{equation}
\langle T_{tt} \rangle + \langle T_{xx} \rangle =  \frac{F(z)}{2 \pi z^2} \int dx x [ (K_0(x) + x K_0'(x))^2 + K_0(x)^2] = \frac{F(z)}{3\pi z^2}, 
\end{equation}
\begin{equation}
\langle T_{zz} \rangle = - \frac{F(z)}{4\pi z^2} \int dx x [ (K_0(x) + x K_0'(x))^2 - x^2 K_0(x)^2 - K_0(x)^2]  =  \frac{F(z)}{6 \pi z^2}. 
\end{equation}
Notice we evaluated these at the BF bound, i.e. $m^2=-1$ in units where the radius of AdS${}_3$ is one $(L=1)$.

\subsubsection{Back-reaction on the metric}

For this Lorentz-invariant case, we can take the metric to be of the form 
\begin{equation}
ds^2 = \frac{dt^2 + dx^2 + g(z) dz^2}{z^2}.
\end{equation}
This is not in the usual Fefferman-Graham gauge, but it will prove to be convenient for our back-reaction calculation. In these coordinates $z$ is more simply related to field theory scale, and the function $g(z)$ directly encodes changes in the AdS scale, corresponding to the renormalisation of the central charge of the dual field theory resulting from the one-loop stress tensor obtained above. 

We will compute the central charge as a function of scale by calculating the Ricci scalar as a function of $z$.
We can do this because the Ricci scalar gives the local effective value of the cosmological constant (or, equivalently, the radially varying space-time length scale $L(z)$) as a function of $z$.   The cosmological constant corresponds to the effective central charge in the dual field theory (see (\ref{Landc})). 

In the interior of the spacetime, for $z \gg z_0$, we can work with a linearised approximation, as the one-loop stress tensor is going to zero. Setting $g= 1 + \delta g$, the linearised Einstein equations give us 
\begin{equation} \label{linR}
R^{(1)} = 6 \delta g - 2 z \delta g' = -2 z^4 \left( \frac{\delta g}{z^3} \right)' = - 2z^2 G_N \langle T_{tt} + T_{xx} + T_{zz} \rangle = -\frac{F(z) G_N}{ \pi}.
\end{equation}
Thus, we see that in the theory with a scalar saturating the BF bound, there is a one-loop running of the central charge. The Ricci scalar is negative (in our units, the background Ricci scalar is $R = -6$), so the correction to the central charge is proportional to $-R^{(1)}$. Thus  the total central charge is 
\begin{equation}
c + \delta c = \frac{3}{2 G_N} + { F(z)  \over 4 \pi}
\label{firstrunningc}
\end{equation}
This decreases towards the infrared as expected, as $F(z)$ decreases as $z \to \infty$ according to the flow equation \eqref{wflow}.    In the IR, the central charge approaches that of the undeformed $F=0$ CFT (which has an AdS representation in terms of the conventional Dirichlet boundary conditions). In the IR region $F(z) \sim 1/\log z$, so the running is logarithmic. 

\subsection{Running of $u$ in the critical flow}
\label{sec:urun}

We now consider the running of $u$ for the momentum dependent double-trace deformation \eqref{Feq}. We consider first the special trajectory where $F_\lambda + F_c = 0$.  It was for this trajectory that we found that the velocity $v$ flows to zero in the IR. The key result is that $u$ approaches a finite value in the IR. 

For this trajectory 
\begin{equation}
\label{eq:Fcri}
F=\frac{\tilde\lambda^2}{v^2}  \frac{k_t^2}{v^2k_x^2+k_t^2}.
\end{equation}
Proceeding as in the momentum-independent case, the renormalised position space Green's function is given by \eqref{eq:green}, but the fact that $F$ and hence $a$ depend separately on $k_x$, $k_t$ implies that the behaviour of the stress tensor components will be different. 

The components of the stress tensor we are interested in are 
\begin{equation}
\label{eq:stress}
\begin{split}
\langle T_{tt} - T_{xx} \rangle &= \langle\partial_t \Phi \partial_t \Phi\rangle - \langle\partial_x \Phi \partial_x \Phi\rangle, \\ 
\langle T_{tt} + T_{xx}\rangle &= - \langle\partial_z \Phi \partial_z \Phi\rangle + \frac{m^2}{z^2} \langle\Phi^2\rangle,  \\
\langle T_{zz}\rangle &= \frac{1}{2} ( \langle\partial_z \Phi \partial_z \Phi\rangle - \langle\partial_t \Phi \partial_t \Phi\rangle - \langle\partial_x \Phi \partial_x \Phi\rangle + \frac{m^2}{z^2} \langle\Phi^2\rangle). 
\end{split}
\end{equation}
The derivatives in this case are given by 
\begin{eqnarray}
  \langle\partial_a\Phi\partial_b\Phi\rangle &=&  \int \frac{d^2k}{4\pi^2 }\,\frac{\left(k_az\,K_0(kz)\right)^2}{A + \log(kz_0) - \frac{v^2}{\tilde \lambda^2} \frac{v^2k_x^2+k_t^2}{k_t^2}}, \label{T1} \\
  \langle\partial_z\Phi\partial_z\Phi\rangle &=& \int \frac{d^2k}{4\pi^2}\,\frac{ \left(K_0(kz) + kz K_0'(kz)\right)^2}{A + \log(kz_0) - \frac{v^2}{\tilde \lambda^2} \frac{v^2k_x^2+k_t^2}{k_t^2}}, \label{T2}
\end{eqnarray}
where $a,b$ run over the boundary coordinates $t,x$. The breaking of Lorentz invariance makes the angular part of the momentum integral non-trivial.

These angular integrals are calculated in appendix \ref{sec:cric} in terms of the quantities 
\begin{equation}
B(k) = A+ \log(kz) + 1/F_\lambda(z) + {\hat v}^2, \quad \hat v^2 = -v^2/F_\lambda = v^2/(\tilde \lambda^2/v^2). 
\end{equation}
Like $a$ in the momentum independent case, $B$ and $\hat v^2$ are independent of the scale we use to evaluate them -- i.e. they are RG invariants.  The result is that the stress tensor contributions in the IR region, where we can make the large $|B|$ approximation,  are 
\begin{equation} \label{tmx}
z^2 \langle T_{tt} - T_{xx} \rangle \approx  - \frac{ \hat v}{6\pi  |B|^{3/2}} \approx - \frac{ \hat v |F_\lambda|^{3/2}}{6 \pi},
\end{equation}
where we used the identity $\int dx x^3\,K_0(x)^2 = \frac{1}{3}$, 
\begin{equation}
\label{eq:tensor1}
z^2 \langle T_{tt} + T_{xx} \rangle \approx  \frac{1}{3\pi |B|} \approx \frac{|F_\lambda|}{3\pi },
\end{equation}
and 
\begin{equation}
\label{eq:tensor2}
z^2 \langle T_{zz} \rangle \approx   \frac{1}{6\pi |B|} \approx \frac{|F_\lambda|}{6\pi }. 
\end{equation}
Thus, the Lorentz-invariant components have a similar form to the momentum-independent case, but with $F_\lambda$ in place of $F$. The  $\langle T_{tt} - T_{xx} \rangle$ component is subleading compared to these in the region of small $F_\lambda$, but as it gives the back-reaction on the velocity, it is our main subject of interest. 

\subsubsection{Backreaction on the metric} 
\label{sec:uback}

As we have broken Lorentz symmetry in this case, we need to take our general ansatz for the metric, writing
\begin{equation} \label{linmetbis}
  ds^2 = \frac{e^{2 \nu(z)} dt^2 + dx^2 + g(z) dz^2}{z^2} \,.
\end{equation}
Then $e^{\nu(z)}$ will parameterize the departure from Lorentz invariance, and give the RG flow for the velocity $u$. Writing $g = 1 + \delta g$, the linearised Einstein equations are  
\begin{equation}
   z^2 \nu'' - z \nu' = - z^2 G_N \langle T_{tt} - T_{xx} \rangle\,,
\label{eq:linear}
\end{equation}
\begin{equation}
- \delta g + \frac{1}{2} z \delta g' = z^2 G_N \langle T_{tt} \rangle,
\end{equation}
and a constraint, 
\begin{equation} \label{econs}
-\delta g  - z \nu' = z^2 G_N \langle T_{zz} \rangle.
\end{equation}
where primes denote derivatives with respect to $z$. Working with the gauge choice \eqref{linmet} rather than the usual Fefferman-Graham coordinates is convenient because it simplifies these linearised equations (fewer derivatives are involved) and will turn out to give a solution where the linearised analysis is valid throughout the IR region.  

If there were no source terms $\langle T_{\mu\nu} \rangle$, these equations would just have the homogeneous solution $\nu_{\text{hom}} = \nu_0 + \nu_1 z^2$, $\delta g_{\text{hom}} = -2 \nu_1 z^2$, where the $z^2$ parts are related by the constraint equation. The constant homogeneous mode $\nu_0$ just corresponds to the freedom to rescale $t$ in the metric \eqref{linmet}. The homogeneous mode $\nu_1$ corresponds to a finite energy density in the dual CFT; it is the linearised version of a black hole solution. Because we are interested in the RG flow in vacuum, we set $\nu_1=0$.  

Consider first the back-reaction of the source in \eqref{eq:linear}. In the region of large $z$, \eqref{tmx} tells us $\langle T_{tt} - T_{xx} \rangle \sim |F_\lambda|^{3/2}$, and the leading order RG equation \eqref{FlRG} tells us that in the IR $F_\lambda \approx 1/\log (z/z_0)$. So  
\begin{equation} 
 z^2 \nu'' - z \nu' \approx   \frac{ G_N \hat v}{6 \pi  \log(z/z_0)^{3/2}}. 
\end{equation}  
Introducing a variable $x = \log(z/z_0)$, the solution to this equation when we set $\nu_1 = 0$ is 
\begin{equation} \label{nusoln}
  \nu = \nu_0 + \frac{\hat v G_N}{3\pi \sqrt{x}} + \ldots. \approx \nu_0 + \frac{\hat v G_N |F_\lambda(z)|^{1/2}}{3\pi}.
  \end{equation}

More formally, to obtain a first-order RG equation from \eqref{eq:linear} the equation for $\nu$ can be written as
\begin{equation}
\bigl( \nu'/z \bigr)'= \frac{ {\hat v} G_N |F_\lambda(z)|^{3/2}}{6 \pi z^3} \, ,
\end{equation}
so we can integrate to write 
\begin{equation}
\nu'(z)= - \bar \nu_1 z + \frac{\hat v G_N}{6 \pi} z \int_{z_0}^{z} du \,  |F_\lambda(u)|^{3/2}/u^3= \\
2 \nu_1 z - \frac{\hat v G_N}{6 \pi} z \int_{z}^\infty du \, |F_\lambda(u)|^{3/2}/u^3
\end{equation}
The $\nu_1$ piece just corresponds to the $z^2$ homogeneous mode in $\nu$. We choose to work at zero temperature (i.e., no background energy density in the CFT and no black hole in the bulk), which corresponds to setting $\nu_1=0$. (Working at finite temperature would be more difficult because we would need to go beyond this linearized approximation in the IR.) This gives us the first-order equation. Recalling that we can  identify $z$ with the scale $\mu$ as $z = \Lambda/\mu$,   and multiplying both sides of the equation by $z$, we can write
\begin{equation}
\partial_{log(\Lambda/\mu)} \nu = -\frac{\hat v G_N}{6\pi} \int_1^\infty dw  \, |F_\lambda(z w)|^{3/2}/w^3 
\end{equation}
in terms of $w= u/z$. Since the running of the coupling $F_\lambda$ is logarithmic, if we neglect the far IR contribution to the integral we can approximate it by taking $F_\lambda(zw) \approx F_\lambda(z)$. Then 
\begin{equation}
\partial_{log(\Lambda/\mu)} \nu = -\frac{\hat v  G_N |F_\lambda(\Lambda/\mu)|^{3/2}}{6 \pi} 
\end{equation}
This has all the properties of an RG equation  in the sense that it depends only on the couplings at the scale - ie there is no remnant of the scale $\Lambda$. 
In the IR, $F(z) \sim  1/\log z$, so if we denote $x=\log(\Lambda/\mu)$ we obtain the equation
\begin{equation}
\partial_x\nu \propto -{1/x^{3/2}},
\end{equation}
which reproduces the solution \eqref{nusoln}.    

Thus, the solution for $\nu$ will approach some constant value in the IR. To determine the constant $\nu_0$, we should fix $\nu$ at some cutoff scale. We could take this cutoff scale somewhere in the IR where the above expression is valid, and use the velocity there to parametrize the different flows. More physically, we should  relate this to the UV velocity, but that is more difficult as we can not control the evolution outside of the IR region.

As discussed above, the velocity $u$ is the ratio of the $g_{tt}$ and $g_{xx}$ components in the metric, and so is given by $e^{2 \nu(z)}$ at the scale $z$.     So the fact that $\nu$ remains finite as we flow to the IR indicates that the velocity is not going to zero. In fact, since we remain in the linearised regime, there is just some small (order $\hbar$) correction to the velocity. Since $\nu$ is decreasing as $x$ increases,  we see that the velocity always gets smaller in the IR, as in the perturbative field theory calculation in section \ref{c2gen}.  On general grounds we might expect that velocities will decrease towards the IR because high energy excitations should effectively provide a ``drag'' for low energy excitations. 
   
The renormalisation of the central charge can be determined as in the previous momentum independent case. We now have that the linearised Ricci scalar is 
\begin{equation} \label{linR1}
R^{(1)} = 6 \delta g - 2 z \delta g'+ 4z\nu^\prime - 2z^2\nu^{\prime\prime} = -2 z^4 \left( \frac{\delta g}{z^3} + \frac {\nu'}{z^2} \right)' , 
\end{equation}
and this is still given by Einstein's equations in terms of the trace of the one-loop stress tensor, 
\begin{equation} \label{linR2}
R^{(1)} = - 2z^2 G_N \langle T_{xx} + T_{yy} + T_{zz} \rangle = -\frac{G_N |F_\lambda|}{ \pi},
\end{equation}
Thus the total central charge is
\begin{equation}
c + \delta c = \frac{3}{2G_N} + \frac{ |F_\lambda|}{4 \pi}
= \frac{3}{2 G_N} + \frac{F_c}{4 \pi}
\label{secondrunningc}
\end{equation}
where we used the fact that $F_c = -F_\lambda$ for the critical flow.    So the only change compared to \eqref{firstrunningc}  is that \eqref{secondrunningc} is expressed in terms of $F_c = |F_\lambda|$.   Below we will consider the general case where $F_c + F_\lambda \neq 0$ and we will find that the central charge is still expressed in terms of $F_c$.     

The one-loop correction to the central charge decreases along the flow, and goes to zero in the IR,  consistent with the idea that the theory flows back to the undeformed fixed point in the IR (i.e. $F=0$ corresponding to an AdS description with Dirichlet boundary conditions).    Interestingly, the running central charge has a universal expression here in terms of the trace of the bulk stress tensor, and can be related to a total derivative in the coordinate system we have adopted here. The significance of this total derivative form is not entirely clear. 
 
If we wanted to use Fefferman-Graham coordinates, we would need to define a coordinate $z = \bar z (1 + \epsilon(\bar z))$ with $ g(z) dz^2 = \frac{d \bar z^2}{\bar z}^2$, that is to linear order 
 \begin{equation}
\frac{(1 + \delta g + 2 \epsilon + 2 z \epsilon')}{1 + 2\epsilon} = 1, 
\end{equation}
Now in the IR region $\delta g \sim \frac{1}{\log z}$, so 
  \begin{equation}
\epsilon \sim  \int \frac{dz}{z \log z} \sim \log (\log z).  
\end{equation}
Thus, the coordinate transformation to Fefferman-Graham coordinates has a slow divergence in the IR; this is why it was useful to work in the coordinate system we have adopted. 

\subsection{More general flows}

The formalism developed in the previous subsection to compute the RG flow of $u$ for the critical flow $\left(F_\lambda + F_c =0\right)$ can equally be applied for the other stable flows, i.e. those with $F_\lambda + F_c >0$. We find that for these flows the velocity $u$ will still be decreasing in the IR, to some finite value, but more rapidly than in the critical flow. 

The calculation proceeds similarly: the background metric ansatz will still be given by \eqref{linmetbis} and the form of the linearised Einstein's equations will remain as in \eqref{eq:linear}-\eqref{econs}. The expectations values $\langle T_{ab} \rangle$ appearing on the right hand side of these equations are still given by \eqref{eq:stress}.  

For flows with $F_\lambda + F_c >0$,  the boundary condition $\Phi_- = F\Phi_+$ involves a double-trace coefficient
\begin{equation}
\label{eq:Fgen}
F = \frac{ (F_c + F_\lambda) v^2 k_x^2 + F_c k_t^2}{v^2 k_x^2 + k_t^2}\,.
\end{equation}
This determines the $\Phi$ Green's function \eqref{eq:green}, through the RG invariant $a$ \eqref{a}. To make its scale independence more manifest, we introduce the RG invariants (i.e. z-independent quantities derived from \eqref{FlRG}) 
\begin{equation}
\hat F = \frac{(F_c + F_\lambda) v^2}{F_c}, \quad \bar v^2 = - \frac{v^2 F_\lambda}{F_c^2}. 
\end{equation}
We can then write 
\begin{equation}
\frac{1}{F} = \frac{1}{F_c} + \frac{\bar v^2 k_x^2}{\hat F k_x^2 + k_t^2} = \frac{1}{F_c} - \frac{\bar v^2}{(1- \hat F)} +  \frac{\bar v^2}{(1- \hat F)} \frac{k_x^2 + k_t^2}{\hat F k_x^2 + k_t^2}. 
\end{equation}
Thus 
\begin{equation}
a = A + \log (kz_0) - \frac{1}{F(z_0)} = B - \hat v^2\frac{k_x^2 + k_t^2}{\hat F k_x^2 + k_t^2}\,,
\end{equation}
where for ease of notation, two modified scale invariants were introduced
\begin{equation}
B = A + \log(k z_0) - \frac{1}{F_c(z_0)} + \hat v^2\quad \text{and} \quad \hat v^2 = \frac{\bar v^2}{(1- \hat F)}\,.
\end{equation}

The calculation of $\langle T_{\mu\nu} \rangle$ proceeds as in subsection~\ref{sec:urun} and it is discussed in appendix \ref{sec:b2}. The $\langle T_{zz}\rangle$ or $\langle T_{tt} + T_{xx} \rangle$ components are dominated by the pole at the origin 
\begin{equation}
  z^2\langle T_{tt} + T_{xx} \rangle\approx \frac{1}{3\pi |B|}\approx \frac{F_c}{3\pi}\,, \quad\quad
  z^2\langle T_{zz}\rangle \approx \frac{1}{6\pi |B|}\approx \frac{F_c}{6\pi}\,. 
\end{equation}
Thus, their values are the same as for the critical flow \eqref{eq:tensor1}-\eqref{eq:tensor2}, except for $|F_\lambda|$ being replaced by $F_c$. The $\langle T_{tt} - T_{xx}\rangle$ component now has comparable contributions from all poles. These are computed in \eqref{eq:btwo} and give rise to
\begin{equation}
\label{eq:ns}
z^2 \langle T_{tt} - T_{xx}\rangle \approx \frac{1-\sqrt{\hat F}}{6\pi \sqrt{\hat F}(1+\sqrt{\hat F})} \frac{\hat v^2}{B^2}\approx  \frac{1-\sqrt{\hat F}}{6\pi \sqrt{\hat F}(1+\sqrt{\hat F})} \hat v^2\,F_c^2 \,.
\end{equation}  
This expectation value is more suppressed in the IR than the one in \eqref{tmx} for the critical flow. Thus, the velocity $\nu$ will decrease more quickly along these flows. 

Indeed, consider \eqref{eq:linear} with the source \eqref{eq:ns}. As in section~\ref{sec:uback}, we will ignore the homogeneous mode associated with thermal effects. In the deep IR, we know from \eqref{eq:fc} that $F_c \approx 1/x$ with $x = \log(z/z_0)$.
Thus, the solution for the linearised running of the velocity in this IR regime will then be 
\begin{equation}
\nu \approx \nu_0 +\frac{G_N}{6\pi}\frac{1-\sqrt{\hat F}}{\sqrt{\hat F}(1+\sqrt{\hat F})} \frac{\hat v^2}{x} + \ldots \approx 
\nu_0 +\frac{G_N}{6\pi}\frac{1-\sqrt{\hat F}}{\sqrt{\hat F}(1+\sqrt{\hat F})} \hat v^2\,F_c\,. 
\end{equation}
This again decreases to some constant value $\nu_0$, but it does so more rapidly than for the critical flow. 

Since the contributions to the trace of the stress tensor have the same form as for the critical flow, the running of the central charge will be given by the same expressions (\ref{linR2}) in terms of $F_c$. 
It is interesting to note that independent of the value of $F_\lambda$, the running of the central charge is determined by the momentum-independent double-trace deformation $F_c$.

\section{Summary}
\label{sec:sum}

We have suggested extensions of the zig-zag phase transition model in \cite{sitte}.  The latter consists of a free boson (realizing a Luttinger liquid) coupled to a Majorana fermion. Phase space is parameterised by the fermion velocity $u$, the Luttinger velocity $v$ and their interaction. The interaction breaks Lorentz invariance but RG flow restores it in the IR where both velocities become equal and quenched. In the model of \cite{sitte} there is a critical line at $u=v$.

In this work, we replaced the Majorana fermion with a more general CFT, including theories with a large central charge and an AdS dual.  First, we studied the $k$ Majorana fermion extension. This theory admits an additional marginal operator parameterized by a new coupling $f_c$, and shows qualitative differences with the $k=1$ case in \cite{sitte}: the velocities $u$ and $v$ decrease towards the IR, but for $k>1$ the velocity $u$ remains finite even in the IR, and in general does not approach $v$. For large $k$, the running of $u$ is parametrically suppressed relative to the running of $v$.  For $k>1$ the critical line in the RG flows is at $f_c = \lambda^2/v^2$ (this bears no relation to the $u=v$ critical line in the $k=1$ case).   

We solved the RG equations analytically in some special cases and numerically for generic values of $k$. For $k=2$ the equations are simple enough to be solved analytically with some assumptions. In the super-critical case ($f_c > \lambda^2/v^2$), $v$ runs to a finite value in the IR,  while in the critical case it goes to zero logarithmically.  In the subcritical case ($f_c < \lambda^2/v^2$), $v$ vanishes at a finite scale.   In all $k=2$ cases, $f_c$ remains finite at all scales.  For $k>2$ we solve the RG flow numerically and found a similar  structure.  In the super-critical case  $v$ remains finite in the IR,  but $f_c$ now goes to zero.    In the critical case both $v$ and $f_c$ flow to zero logarithmically. In the subcritical case $v$ again goes to zero at a finite scale.

Given the simplifications of the large $k$ fermion model, we considered a further generalisation in which we replaced the CFT describing the $k$ Majorana fermions with an arbitrary large central charge CFT containing a dimension $(1/2,1/2)$ operator $\mathcal O$ coupled to the Luttinger theory and allowing a marginally irrelevant deformation $\mathcal O^2$. We then used the AdS/CFT correspondence to analyze these more general scenarios. The qualitative structure of the bulk RG flows reproduces the field theory behavior for $k>2$. In the large central charge limit, the running of $u$ is subleading; we studied it by including gravitational back-reaction in the holographic calculation.  Our technique for studying the problem involves integrating out the Luttinger liquid entirely and then studying the resulting deformed conformal theory.  This allowed us to avoid the question of how the degrees of freedom of the Luttinger liquid were specifically realized in an AdS model.  This could have been alternatively treated in the semi-holographic approach of \cite{Faulkner:2010tq}.     Note that the AdS analysis computes the RG behavior for a a generic large central charge CFT,  assuming only the presence of a dimension one operator that couples to the current, and effectively sums up all the planar Feynman diagrams,  thus going well beyond the one-loop analysis of \cite{sitte}.

Some of our proposed generalizations can be experimentally realized.  Consider, for example, a three dimensional material with a potential in two transverse directions such that two transverse oscillations become massless at the same value of the chemical potential.  This will realize our $k=2$ model.

\section*{Acknowledgements}
MB would like to thank Ehud Altman for enlightening and useful discussion at early stages of this project. The work of JS was partially supported by the Engineering and Physical Sciences Research Council (EPSRC) [grant number EP/G007985/1] and the Science and Technology Facilities Council (STFC) [grant number ST/J000329/1].  VB was supported by DOE grant DE-FG02-05ER-41367 and by the Fondation Pierre-Gilles de Gennes.  SFR was supported by STFC and the Institut Lagrange de Paris. MB was supported by the Israeli science foundation centers of excellence program, the German-Israeli foundation for scientific research and development and the Minerva foundation.   VB and MB were supported in early stages of this project by the US-Israel Bi-National Science Foundation.


\appendix

\section{T-duality and the appearance of ${\cal O}^2$}
\label{tduality}

We want to understand the transformation between the descriptions of the Luttinger liquid in terms of the two dual variables $\theta$ and $\phi$ when we include the coupling to $\mathcal O$. We start from the Luttinger liquid Hamiltonian \eqref{hamv1} and the canonical commutation relations \begin{equation}
\left[\frac{1}{\pi}\partial_x\phi (x),\,\theta (x^\prime)\right] = -i \delta(x-x^\prime).
\end{equation}
When we have only a Luttinger liquid, then the derivatives are dual, $\partial_{t,x} \theta\propto \partial_{x,t}\phi$. As a result, for $k=1$, the interaction Lagrangians $\int \partial_t\theta \, {\cal O}$ and $\int \partial_x\phi  \, {\cal O}$ are related. However we will see that for $k\ge 2$ the two interaction Lagrangians are not  the same, but rather they differ by the addition of the operator ${\cal O}^2$ with a specific coefficient. 

It is simplest to see this in the path integral formulation of the Luttinger liquid. There are two equivalent (Minkowskian) Lagrangian descriptions of the Luttinger theory, one using only $\theta$ and one using only $\phi$,
\begin{equation}\label{lagthet}
{\cal L}_\theta={K \over 2\pi v} \int dt dx \bigl( (\partial_t\theta)^2 - v^2 (\partial_x\theta)^2 \bigr) ,
\end{equation}
\begin{equation}\label{lagphi}
{\cal L}_\phi = {1\over 2\pi K v} \int dt dx \bigl( (\partial_t\phi)^2 - v^2 (\partial_x\phi)^2 \bigr).
\end{equation}
We will further rescale the coordinate $x^{new}= x/v$,which sets $v=1$.

We can go from one Lagrangian to the other by starting with the path integral 
\begin{equation}\label{lftlag}
\int D\phi DV_\mu exp \biggl(- {i\over 2\pi} \int dt dx \bigl( K \eta^{\mu\nu} V_\mu V_\nu + 2\epsilon^{\mu\nu}\phi\, \partial_\mu V_\nu \bigr) \biggr)
\end{equation}
where $\eta^{00}=-1,\ \eta^{11}=1$.
This path integral reduces to \eqref{lagthet} if we treat $\phi$ as a Lagrange multiplier and solve the resulting equation of motion by setting $V_\mu=\partial_\mu \theta$, or it reduces to \eqref{lagphi} if we integrate out $V_\mu$ and keep $\phi$ as a dynamical variable.

Suppose that we are now given a Lagrangian of the form
\begin{equation}\label{adefrm}
{\cal L'}_\theta={\cal L}_\theta + (\alpha_t \partial_t \theta-\alpha_x\partial_x \theta) {\cal O} - {\alpha} {\cal O}^2.
\end{equation}
We can replace it by 
\begin{equation}\label{lftdfrm}
-{\cal L}=  \frac{K}{2\pi} \eta^{\mu\nu} V_\mu V_\nu + \frac{1}{\pi} \epsilon^{\mu\nu} \phi\, \partial_\mu V_\nu + \eta^{\mu\nu}\alpha_\mu V_\nu {\cal O} + \alpha {\cal O}^2,
\end{equation}
where as before we reduce to \eqref{adefrm} by integrating out $\phi$ to obtain $V_\mu=\partial_\mu\theta$.
Now we can integrate out $V_\mu$ to obtain the Lagrangian
\begin{equation}
-{\cal L}= {1\over 2\pi K}\eta^{\mu\nu}\partial_\mu\phi\partial_\nu\phi -  \frac{1}{K}\epsilon^{\mu\nu} \alpha_\mu \partial_\nu \phi\, {\cal O} +  \left(\alpha - {2\pi \over K} \eta^{\mu\nu}\alpha_\mu\alpha_\nu\right) {\cal O}^2
\end{equation}
or
\begin{equation}
\alpha_t^{(\phi)}=\frac{1}{K}\alpha_x^{(\theta)}
\end{equation}
\begin{equation}
\alpha_x^{(\phi)}=\frac{1}{K}\alpha_t^{(\theta)}
\end{equation}
\begin{equation}\label{alphshft}
\alpha^{(\phi)}=\alpha^{(\theta)}-\frac{2\pi}{K}\eta^{\mu\nu}\alpha_\mu^{(\theta)}\alpha_\nu^{(\theta)}.
\end{equation}
The last term in \eqref{alphshft} is the shift in the coefficient of $O^2$. In particular, if we start with only $ \lambda^{(\theta)}_t$, then we end up with ${\lambda}^{(\phi)}_x= {\lambda}^{(\theta)}_t/K$, and $\alpha^{(\phi)}=\frac{2\pi}{K} \bigl({\lambda}^{(\theta)}_t\bigr)^2 = 2\pi K ({\lambda}^{(\phi)}_x\bigr)^2$, which agrees with the discussion around  \eqref{dtphi} and \eqref{dttheta}, noting we both set $v=1$ and work in Lorentzian signature in this appendix, while $v$ is generic and the signature is Euclidean in the main body of the paper.

\section{Evaluation of $\langle T_{\mu\nu}\rangle$}
\label{angle}

In this appendix we compute the momentum integrals determining the expectation value of the stress tensor $\langle T_{\mu\nu}\rangle$. The latter are given in \eqref{eq:stress} and are fully determined by \eqref{T1}-\eqref{T2}
\begin{equation}
\begin{split}
  \langle\partial_a\Phi\partial_b\Phi\rangle &=  \int \frac{d^2k}{4\pi^2 a}\,\left(k_az\,K_0(kz)\right)^2 ,  \\
  \langle\partial_z\Phi\partial_z\Phi\rangle &= \int \frac{d^2k}{4\pi^2 a}\,\left(K_0(kz) + kz K_0'(kz)\right)^2,
\end{split}
\end{equation}
where the scale independent $a$ equals
\begin{equation}
  a = A + \log(kz_0) - \frac{1}{F}\,.
\label{eq:newa}
\end{equation}
As discussed in the main text, it is $F$ that controls the kind of double trace deformation (flow) under consideration.
To study these integrals, we apply two transformations. First, we work with polar coordinates in momentum space
\begin{equation}
  k_t=k\sin\theta,\quad \quad k_x=k\cos\theta\,. 
\end{equation} 
Second, we introduce a complex coordinate $w$ mapping the angular $\theta$ integral to the unit circle closed contour in the complex plane
\begin{equation}
  w=e^{i\theta} \quad \Rightarrow \quad \cos\theta = \frac{(w+1/w)}{2}\,, \,\,\, \sin\theta = \frac{(w-1/w)}{2i}\,.
\end{equation}

\subsection{Critical flows}
\label{sec:cric}

First, we discuss these integrals for critical flows, i.e. $F_\lambda + F_c=0$, where $F$ is given in \eqref{eq:Fcri}.
Using the above transformations, we can jointly write $\langle T_{\mu\nu}\rangle$ as
\begin{equation}
\langle T_{\mu\nu}\rangle =  \int \frac{kdk}{4\pi^2} H(kz)\,\left(\frac{1}{i}\oint \frac{dw}{w}\,\frac{(w^2-1)^2}{B(w^2-1)^2 + 4w^2 \hat v^2}\,G(w)\right).
\label{eq:int}
\end{equation}
In writing this expression, we introduced the RG-invariants 
\begin{equation}
B = A+ \log(kz) + 1/F_\lambda(z) + {\hat v}^2, \quad \hat v^2 = -\frac{v^2}{F_\lambda} = \frac{v^2}{\tilde \lambda^2/v^2}. 
\end{equation}
The different relevant components are characterised by
\begin{itemize}
\item  For $\langle T_{tt} - T_{xx}\rangle$, we have $G(w) = - \frac{w^4+1}{2w^2}$ and $H(kz) = (kz K_0(kz))^2$. 
\item For $\langle T_{zz}\rangle$, $G(w)=1$ and
\begin{equation} 
H(kz) = \frac{1}{2} [ (K_0(kz) + kz K_0'(kz))^2 - k^2 z^2 K_0(kz)^2 + m^2 K_0(kz)^2].
\end{equation}
\item For $\langle T_{tt} + T_{xx} \rangle$, we again have $G(w)=1$ and 
\begin{equation} 
H(kz) =  [ -(K_0(kz) + kz K_0'(kz))^2 + m^2 K_0(kz)^2].
\end{equation}
\end{itemize}

Note that as in the momentum independent flows discussed in section~\ref{sec:wup}, far in the infrared, $B \approx 1/F_\lambda(z)$, as $F_\lambda \to 0$ in the IR. In performing the integral, we will always assume that we are in this region of large $B$. As a result the angular integral is approximately independent of $k$, allowing us to compute it separately from the integral over $k$. 

\paragraph{Angular integral:} Consider first the angular integral with $G(w) = 1$. Using the residue theorem, there can only be contributions from the poles in the integrand. There is a manifest simple pole at $w=0$, whose contribution to the unit circle integral is $2\pi/B$. Let us examine the existence of further poles in the second factor :
\begin{equation*}
  B(w^2-1)^2 + 4\hat v^2(w^2-1) + 4\hat v^2 = 0\,.
\end{equation*}
In the limit of large $|B|$, when we are in the deep interior of the spacetime, i.e. in the IR of the RG flow, $|B|\gg \hat v^2$, this equation reduces to
\begin{equation}
  w_{\star\pm}^2 - 1 = \mp \frac{2\hat v}{\sqrt{-B}}\,.
\end{equation}
Notice that only $w_{\star +}$ is inside the contour of integration, giving rise to two simple poles,
\begin{equation}
  w_\pm \sim \pm\left(1-\frac{\hat v}{\sqrt{-B}}\right)\,.
\end{equation}
Using the residue theorem, their contribution to the integral is 
\begin{equation*}
\frac{2\pi i}{i}\left[\frac{1}{w_+}\frac{(w_+^2-1)^2}{B(w_+-w_-)(w_{\star+}^2-w_{\star-}^2)} + \frac{1}{w_-}\frac{(w_-^2-1)^2}{B(w_- -w_+)(w_{\star+}^2-w_{\star-}^2)}\right]
\end{equation*}
In the large $|B|$ limit, where $w_+-w_- \approx 2$, the latter is
\begin{equation*}
  \frac{2\pi}{|B|} \frac{\hat v}{\sqrt{|B|}}
\end{equation*}
up to subleading contributions. Thus, the overall integral is dominated by the pole at the origin, and
\begin{equation}
  \frac{1}{i}\oint \frac{dw}{w}\,\frac{(w^2-1)^2}{B(w^2-1)^2 + 4w^2 \hat v^2} \approx \frac{2\pi}{B}. 
\label{eq:origin}
\end{equation}
To sum up, for the components described by $G(w)=1$, the result has the same form as in the momentum-independent case.
This is because the integral is dominated by the pole at the origin and the angular dependence only makes a subleading contribution to this integral. 

Consider the case with $G(w) = - \frac{w^4+1}{2w^2}$. For the momentum-independent case the corresponding integral vanishes. Thus, the angular dependence must be crucial now. There is still a contribution from the pole at the origin, but the current non-trivial $G(w)$ turns this into a third order pole. To evaluate its contribution, we use Cauchy's integral formula
\begin{equation*}
  f^{(n)}(a) = \frac{n!}{2\pi i} \oint \frac{f(w)}{(w-a)^{n+1}}\,dw\,.
\end{equation*}
This gives a contribution from the pole at the origin of $\frac{4\pi \hat v^2}{B^2}$. The integral still has the same two poles $w_\pm$ identified earlier. Since the new function $G(w)$ is even, we can again conclude the contribution of both poles is equal. In fact since $G(w_\pm) \sim -1$, these poles have the same contribution as before. As a result, the contribution from the origin will now be subleading, allowing us to approximate the integral by
\begin{equation}
  - \frac{1}{i}\oint \frac{dw}{w}\,\frac{(w^2-1)^2}{B(w^2-1)^2 + 4w^2 \hat v^2}\,\frac{w^4+1}{2w^2} \approx -\frac{2\pi}{|B|}\frac{\hat v}{\sqrt{|B|}}\,.
\label{eq:crit-x}
\end{equation}

\subsection{More general flows}
\label{sec:b2}

In this subsection, we compute the integrals determining $\langle T_{\mu\nu} \rangle$ for the more general stable flows satisfying $F_c + F_\lambda > 0$. Using the value of $F$ in \eqref{eq:Fgen} determining $a$ in \eqref{eq:newa}, we can write these expectation values as
\begin{equation}
\langle T_{\mu\nu}\rangle =  \int \frac{kdk}{4\pi^2} H(kz)\,\left(\frac{1}{i}\oint \frac{dw}{w}\,\frac{(w^2-1)^2 - \hat F (w^2 +1)^2 }{B [(w^2-1)^2 - \hat F (w^2 +1)^2] + 4w^2 \hat v^2}\,G(w)\right),
\label{eq:bone}
\end{equation}
where, for ease of notation, the following RG-invariants were introduced
\begin{equation}
B = A + \log(k z_0) - \frac{1}{F_c(z_0)} + \hat v^2\quad \text{and} \quad \hat v^2 = \frac{\bar v^2}{(1- \hat F)}\,,
\end{equation}
as a function of 
\begin{equation}
\hat F = \frac{(F_c + F_\lambda) v^2}{F_c}, \quad \bar v^2 = - \frac{v^2 F_\lambda}{F_c^2},
\end{equation}
but with the same functions $H(kz)$, $G(w)$ as in the critical flow discussion. 

The contributions from the pole at the origin are similar to previously. Its residue equals $2\pi/B$ for $G(w) =1$ and $4\pi \hat v^2/(B^2(1-\hat F))$ for $G(w) = -(w^4+1)/2w^2$. To determine the remaining poles, notice that for stable flows
$\hat F >0$ and in the IR we can again consider the large $|B|$ limit. Assuming $\frac{\hat v^2}{B}$ is small, we can expand the denominator in \eqref{eq:bone} to linear order around the zeroes in its numerator. These equal 
\begin{equation}
w_{0\pm}^2 = \frac{(1 \pm \sqrt{ \hat F})^2}{(1 - \hat F)}.
\end{equation}
Notice it is only the lower sign $w_{0-}$ that lies inside the unit circle. Solving for the poles\footnote{This expression does not have a smooth limit as $\hat F \to 0$ because we implicitly assumed the $1/B$ term is small compared to $\hat F$. At fixed $B$, there should actually be some crossover between this $1/B\sqrt{\hat F}$ behaviour and the previous $1/\sqrt{B}$ as $\hat F$ decreases.} 
\begin{equation}
w_{-}^2 \approx w_{0-}^2 \left( 1 + \frac{\hat v^2}{B \sqrt{\hat F}} \right).
\end{equation}
As for the critical flows, there are two poles $\pm w_-$ inside the unit circle whose residues add up to
\begin{equation}
2\pi \frac{G(w_{0-})}{\sqrt{\hat F}} \frac{\hat v^2}{B^2} . 
\end{equation}
Thus, for $G(w)=1$, this gives a subleading contribution, whereas for $G(w)=-(w^4+1)/2w^2$, it is of the same order as the contribution from the pole at the origin.

To sum up, since $\langle T_{zz}\rangle$ and $\langle T_{tt} + T_{xx}\rangle$ have $G(w)=1$, their dominant contribution is from the pole at the origin. The latter is given by \eqref{eq:origin}, as for the critical flow. On the other hand, $\langle T_{tt} - T_{xx}\rangle$ has $G(w)=-(w^4+1)/2w^2$ and the value of the angular integral equals
\begin{equation}
  \frac{1}{i}\oint \frac{dw}{w}\,\frac{(w^2-1)^2 - \hat F (w^2 +1)^2 }{B [(w^2-1)^2 - \hat F (w^2 +1)^2] + 4w^2 \hat v^2}\,G(w) \approx -2\pi \frac{1-\sqrt{\hat F}}{\sqrt{\hat F}(1+\sqrt{\hat F})} \frac{\hat v^2}{B^2}\,.
\label{eq:btwo}
\end{equation}

\end{document}